\documentclass[]{aa}

\usepackage[varg]{txfonts}
\usepackage{graphicx}
\begin{document}

\title{Signs of outflow feedback from a nearby young stellar object on the protostellar envelope around HL~Tau}
\author{Hsi-Wei Yen\inst{\ref{inst1},\ref{inst2}}, Shigehisa Takakuwa\inst{\ref{inst3},\ref{inst2}}, Pin-Gao Gu\inst{\ref{inst2}}, Naomi Hirano\inst{\ref{inst2}}, Chin-Fei Lee\inst{\ref{inst2}}, Hauyu Baobab Liu\inst{\ref{inst1},\ref{inst2}}, Sheng-Yuan Liu\inst{\ref{inst2}}, Chun-Ju Wu\inst{\ref{inst4},\ref{inst2}}}
\institute{European Southern Observatory (ESO), Karl-Schwarzschild-Str. 2, D-85748 Garching, Germany\label{inst1}
\and
Academia Sinica Institute of Astronomy and Astrophysics, 11F of Astro-Math Bldg, 1, Sec. 4, Roosevelt Rd, Taipei 10617, Taiwan; \email{hwyen@asiaa.sinica.edu.tw}\label{inst2}
\and 
Department of Physics and Astronomy, Graduate School of Science and Engineering, Kagoshima University, 1-21-35 Korimoto, Kagoshima, Kagoshima 890-0065, Japan\label{inst3}
\and
Department of Physics, National Taiwan University, No. 1, Sec. 4, Roosevelt Road, Taipei 106, Taiwan\label{inst4}
}
\date{Received date / Accepted date}

\abstract
{}
{\object{HL~Tau} is a Class I--II protostar embedded in an infalling and rotating envelope and possibly associated with a planet forming disk, 
and it is co-located in a 0.1 pc molecular cloud with two nearby young stellar objects with projected distance of $\sim$20\arcsec--30\arcsec (2800--4200 au) to \object{HL~Tau}.
Our observations with the Atacama Large Millimeter/Submillimeter Array (ALMA) revealed two arc-like structures on a 1000 au scale connected to the disk,  
and their kinematics could not be explained with any conventional model of infalling and rotational motions. 
In this work, we investigate the nature of these arc-like structures connected to the \object{HL~Tau} disk.}
{We conducted new observations in the $^{13}$CO and C$^{18}$O (3--2; 2--1) lines with the James Clerk Maxwell Telescope and the IRAM 30m telescope, and obtained the data with the 7-m array of the Atacama Compact Array (ACA). 
With the single-dish, ACA, and ALMA data, we analyzed the gas motions on both 0.1 pc and 1000 au scales in the \object{HL~Tau} region. 
We constructed new kinematical models of an infalling and rotating envelope with the consideration of relative motion between \object{HL~Tau} and the envelope.}
{By including the relative motion between \object{HL~Tau} and its protostellar envelope, 
our kinematical model can explain the observed velocity features in the arc-like structures. 
The morphologies of the arc-like structures can also be explained with an asymmetric initial density distribution in our model envelope.
In addition, our single-dish results support that \object{HL~Tau} is located at the edge of a large-scale (0.1 pc) expanding shell driven by the wind or outflow from \object{XZ~Tau}, as suggested in the literature.
The estimated expanding velocity of the shell is comparable to the relative velocity between \object{HL~Tau} and its envelope in our kinematical model. 
These results hints that the large-scale expanding motion likely impacts the protostellar envelope around \object{HL~Tau} and affects its gas kinematics. 
We found that the mass infalling rate from the envelope onto the \object{HL~Tau} disk can be decreased by a factor of two due to this impact by the large-scale expanding shell.
}
{}


\titlerunning{Outflow feedback on the protostellar envelope around HL~Tau}
\authorrunning{H.-W. Yen et al.}

\maketitle

\section{Introduction}
\object{HL~Tau} is a Class I--II protostar \citep{Men'shchikov99, Motte01, Furlan08} located in the Taurus star-forming region at a distance of 140 pc  \citep{Kenyon94, Loinard13, Galli18}. 
\object{HL~Tau} is surrounded by a protostellar envelope with a size of 2000 au \citep{Hayashi93}, 
and the protostellar envelope is embedded in a molecular cloud with a length of 0.05 pc elongated along the northwest--southeast direction \citep{Welch00}. 
The protostellar envelope around \object{HL~Tau}  exhibits signs of infalling and rotational motions on a 1000 au scale as observed in the $^{13}$CO (1--0) emission at an angular resolution of 5$\arcsec$ \citep{Hayashi93}. 
Nevertheless, the overall velocity structures in the protostellar envelope are complex \citep{Cabrit96, Welch00, Wu18}. 
Bipolar molecular outflows and high-velocity (60--160 km s$^{-1}$) optical and infrared jets associated with \object{HL~Tau} are observed \citep{Monin96, Takami07, Hayashi09, Anglada07, Movsessian12, Lumbreras14, ALMA15, Klaassen16}. 
The presence of the infalling envelope and the outflow activities suggests that \object{HL~Tau} is still in the main accretion phase and is less evolved compared to other Class II protostars. 

Observations in the (sub-)millimeter continuum with the Atacama Large Millimeter/Submillimeter Array (ALMA) revealed a series of rings and gaps in the circumstellar disk around \object{HL~Tau} \citep{ALMA15, Akiyama16}.
The presence of such rings and gaps can be a signpost of gas giant planets embedded in the disk \citep[e.g.,][]{Dong15, Dong16, Kanagawa15, Kanagawa16, Dong17}.
The imaging observations in the $L^\prime$ band with the Large Binocular Telescope Interferometer did not find point sources in the disk, 
and the upper limit on the mass of the putative planets is estimated to be 10--15 $M_{\rm jup}$ \citep{Testi15}. 
Nevertheless, this upper limit is more than ten times higher than the estimated required mass for a planet to carve the observed gaps in the disk around \object{HL~Tau}  \citep{Kanagawa15, Akiyama16, Jin16, Yen16}. 
These observational results make \object{HL~Tau} one of the youngest candidates of on-going planet formation. 

When there is material infalling onto a protoplanetary disk from its surrounding envelope, 
the disk mass increases, and the disk can become gravitational unstable \citep{Machida10, Vorobyov10}.
That may lead to the formation of progenitors of planets or mass ejection from the disk \citep{Vorobyov11, Vorobyov16, Zhu12}.  
Infalling material onto protoplanetary disks can also produce accretion shock and affect the chemistry in the disks \citep{Visser09, Visser11}.
In order to study the gas kinematics of the protostellar envelope around \object{HL~Tau} and its influence on the \object{HL~Tau} disk, 
we conducted ALMA observations in the $^{13}$CO (2--1) and C$^{18}$O (2--1) lines at an angular resolution of 0\farcs8  \citep{Yen17}.

Our previous ALMA observations revealed two arc-like structures with sizes of 1000 au and 2000 au and masses of $3 \times 10^{-3}$ $M_\odot$ connected to the \object{HL~Tau} disk in the protostellar envelope. 
One is blueshifted and stretches from the northeastern part of the disk toward the northwestern direction, and the other is redshifted and stretches from the southwestern part of the disk toward the southeastern direction.
The kinematics of the arc-like structures could not be explained with the conventional models of an infalling and rotating envelope. 
In addition, the observed velocity in the blueshifted arc-like structure is higher than the expected free-fall velocity in \object{HL~Tau}. 
The origins of the arc-like structures are unclear. 
Two possibilities are suggested in \citet{Yen17}, infalling flows under the external compression or outflowing gas caused by gravitational instabilities in the disk. 

To investigate the origins of the arc-like structures connected to the \object{HL~Tau} disk, 
we have conducted observations in the $^{13}$CO (3--2) and C$^{18}$O (3--2) lines with the James Clerk Maxwell Telescope (JCMT) and in the $^{13}$CO (2--1) and C$^{18}$O (2--1) lines with the IRAM 30m telescope and mapped the kinematics of the large-scale molecular cloud associated with \object{HL~Tau}. 
In addition, we have obtained the data of the $^{13}$CO (2--1) and C$^{18}$O (2--1) lines with the 7-m array of the Atacama Compact Array (ACA), which is a part of our ALMA project. 
With the IRAM 30m, ACA, and ALMA data, 
we generated combined images.
In this paper, we present the single-dish and combined images, and we discuss the physical conditions and kinematics on 0.1 pc and 1000 au scales.  
To explain the observed velocity features in the arc-like structures, 
we construced a new kinematical model for the protostellar envelope around \object{HL~Tau} with the consideration of relative motion between \object{HL~Tau} and the envelope. 
We discuss the possible origin of this relative motion and its effects on the dynamics and evolution of the protostellar envelope around \object{HL~Tau}.

\section{Observations}
The fields of view, the velocity and angular resolutions, and the noise levels of all the observations are summarized in Table \ref{obsum}. 
In this paper, the intensity of the single-dish data is presented in units of main beam brightness temperature. 

\begin{table*}
\caption{Summary of images}\label{obsum}
\centering
\begin{tabular}{cccccc}
\hline\hline
Telescopes & Line & Field of View & Velocity resolution & Angular resolution & Noise per channel  \\ 
\hline 
JCMT & $^{13}$CO (3--2) & $2\arcmin \times 2\arcmin$ & 0.06 km s$^{-1}$ & 15\farcs3 & 0.44 K \\
JCMT & C$^{18}$O (3--2) & $2\arcmin \times 2\arcmin$ & 0.06 km s$^{-1}$ & 15\farcs2 & 0.41 K \\
IRAM 30m & $^{13}$CO (2--1) & $2\arcmin \times 2\arcmin$ & 0.07 km s$^{-1}$ & 11\farcs8 & 0.24 K \\
IRAM 30m & C$^{18}$O (2--1) & $2\arcmin \times 2\arcmin$ & 0.07 km s$^{-1}$ & 11\farcs8 & 0.26 K \\
30m+ACA+ALMA & $^{13}$CO (2--1) & $40\arcsec \times 40\arcsec$ & 0.2 km s$^{-1}$ & $1\arcsec \times 0\farcs9$ (151\degr) & 7.7 mJy Beam$^{-1}$\\
30m+ACA+ALMA & C$^{18}$O (2--1) & $40\arcsec \times 40\arcsec$ & 0.34 km s$^{-1}$ & $1\arcsec \times 0\farcs9$ (160\degr) & 5.5 mJy Beam$^{-1}$\\
\hline
\end{tabular}
\end{table*}

\subsection{JCMT observations}
Our observations in the $^{13}$CO (3--2; 330.587960 GHz) and C$^{18}$O (3--2; 329.330545 GHz) lines toward \object{HL~Tau} with JCMT were conducted on January 14, 17, 29, and 30, February 5, and August 1 in 2017. 
The observations were conducted in the grade 3 weather with 225 GHz opacity ranging from 0.08 to 0.12. 
The on-source observing time on each day ranges from 0.3 to 1.4 hours, and the total on-source time is 4.4 hours. 
The Heterodyne Array Receiver Program (HARP) was adopted. 
HARP is an array receiver with 4 $\times$ 4 detectors, 
which are named as receptor H00, H01 to H15.   
The main beam efficiency of HARP is 0.64. 
The receptors, H13 and H14, were not operational. 
The data taken with the receptors H04 on all the observing days, H08 on January 14, and H03 on February 5 were not usable, 
and they were excluded in our data reduction. 
The observations were conducted with the Jiggle-Position Switch mode, which covers an area of $2\arcmin \times 2\arcmin$ centered at the position of \object{HL~Tau}. 
The spectrometer ACSIS was adopted as the backend and was configured to have two spectral windows in the 250 MHz mode to observe the $^{13}$CO (3--2) and C$^{18}$O (3--2) lines simultaneously. 
In each spectral window, the usable bandwidth is 220 MHz, and the channel number and width are 4096 and 61 kHz, respectively.
That corresponds to a velocity resolution of 0.06 km s$^{-1}$ at the frequencies of the $^{13}$CO (3--2) and C$^{18}$O (3--2) lines. 
The data were reduced with Starlink \citep{Currie14} and the ORAC-DR pipeline \citep{Jenness15}, and the recipe of ``reduce\_science\_narrowline'' was adopted. 
This recipe is designed for data with one or more narrow ($<$8 km s$^{-1}$) emission lines in the observed frequency band. 
In particular, it smooths data by 5 pixels $\times$ 5 pixels $\times$ 10 channels to find regions for fitting and subtracting baselines.
The detailed process is described at \url{http://www.starlink.ac.uk/devdocs/sun260.htx/sun260.html}.
The achieved noise level in main beam temperature is 0.4 K at a velocity resolution of 0.06 km s$^{-1}$. 

\subsection{IRAM 30m observations}
Our observations toward \object{HL~Tau} with the IRAM 30m telescope were conducted on March 28 and 29 in 2017. 
During the observations, the 225 GHz opacity ranged from 0.08 to 0.15. 
The total on-source time is 4.7 hours.
The observations were conducted with the EMIR receiver. 
On-the-fly mapping was performed to cover an area of $2\arcmin \times 2\arcmin$ centered at the position of \object{HL~Tau}.
Dual bands, E90 and E230,  at 3 mm and 1.3 mm with dual polarizations were observed simultaneously. 
The spectrometer FTS was adopted as the backend and was configured in the narrow mode, resulting in a bandwidth of 1.8 GHz and a spectral resolution of 50 kHz for each frequency band and each polarization. 
The observed frequency ranges covered several molecular lines, including HCO$^+$ (1--0), H$^{13}$CO$^+$ (1--0), HCN (1--0), and SiO (2--1) in the E90 band and $^{13}$CO (2--1), C$^{18}$O (2--1), DCN (3--2), SiO (5--4), and SO (5$_6$--4$_5$) in the E230 band. 
In this paper, we present the results of the $^{13}$CO (2--1; 220.398684 GHz) and C$^{18}$O (2--1; 219.560358 GHz) lines. 
The main beam efficiency of the IRAM 30m telescope at the frequency of 220 GHz is 0.61. 
The achieved velocity resolution at the frequencies of the $^{13}$CO (2--1) and C$^{18}$O (2--1) lines is 0.07 km s$^{-1}$.
The data were reduced with CLASS (\url{http://www.iram.fr/IRAMFR/GILDAS}). 
The achieved noise level in main beam temperature is 0.2--0.3 K at a velocity resolution of 0.07 km s$^{-1}$ in the E230 band. 

\subsection{ACA 7-m array observations}
Our observation toward \object{HL~Tau} with the ACA 7-m array is a part of our ALMA project, 2015.1.00551.S. 
The observations were conducted on January 17, June 16, and September 8 in 2016.
A 7-pointing mosaic pattern with the separation between each pointing of 20$\arcsec$ was adopted, 
and the mapping area is larger than the field of view of the ALMA observations with the 12-m array \citep{Yen17}. 
The total on-source time per pointing is 20 minutes. 
On January 17, J0510+1800 (3.4 Jy at 220 GHz) was observed as flux and bandpass calibrators, and J0431+1731 (1.2--1.3 Jy at 220 GHz) as a phase calibrator. 
On June 16 and September 9, the flux calibrators are Uranus and J0522$-$3627 (3.4 Jy at 220 GHz), respectively, and J0522$-$3627 was also observed as a bandpass calibrator. 
On these two days, J0510+1800 was observed as a phase calibrator (2.1 Jy on June 16 and 1.4 Jy on September 9 at 220 GHz). 
The spectral setup of our ACA observations is the same as that of our observations with the 12-m array, which contains five spectral windows.
Two spectral windows, each with a bandwidth of 2 GHz, were assigned to the 1.3 mm continuum. 
One spectral window with a bandwidth of 468.8 MHz (638 km s$^{-1}$) and a channel width of 122 kHz (0.17 km s$^{-1}$) was set to the $^{13}$CO (2--1) line, one with a bandwidth of 234.4 MHz (320 km s$^{-1}$) and a channel width of 244 kHz (0.33 km s$^{-1}$) to the C$^{18}$O (2--1) line, and one with a bandwidth of 234.4 MHz (319 km s$^{-1}$) and a channel width of 488 kHz (0.67 km s$^{-1}$) to the SO (5$_6$--4$_5$) line.
The data were reduced with the pipeline in Common Astronomy Software Applications \citep[CASA;][]{McMullin07} of version 4.7.0-1. 
The ACA 7-m array data were combined with the ALMA 12-m array data to generate images.

\subsection{Combined maps}
We generated combined images of the $^{13}$CO (2--1) and C$^{18}$O (2--1) emission in \object{HL~Tau} with the IRAM 30m, ACA, and ALMA data using CASA of version 4.7.0. 
The details of our ALMA observations were described in \citet{Yen17}.
We first adopted {\it imagermode} of ``mosaic'' and generated interferometric images with the ACA and ALMA data together using the CASA task, {\it clean}. 
These interferometric images of the $^{13}$CO (2--1) and C$^{18}$O (2--1) emission were generated with the natural weighting at velocity resolutions of 0.2 km s$^{-1}$ and 0.34 km s$^{-1}$, respectively. 
Multi-scale clean with the scale sizes of 0, 2$\arcsec$, 5$\arcsec$, and 10$\arcsec$ was performed. 
Then, we re-grided the IRAM 30m images to have the same velocity axis as the interferometric images and multiplied them by the primary beam response of the mosaic observations. 
Finally, we combined the interferometric and IRAM 30m images to generate the combined images using the CASA task, {\it feather}, and its visual interface, {\it casafeather}.
The final resolutions of the combined images are 1$\arcsec$, and the noise levels are 6 and 8 mJy beam$^{-1}$ in the $^{13}$CO (2--1) and C$^{18}$O (2--1) images, respectively (Table \ref{obsum}).
We note that the images from our mosaicking observations with the 7-m array are affected by the imaging issues of CASA\footnote{\url{http://library.nrao.edu/public/memos/naasc/NAASC_117.pdf}}, and the fluxes in our mosaic map are overestimated by 10\%. 
Nevertheless, in this work, we analyze the intensity distributions in the velocity channel maps, 
and our analysis does not depend on the flux level. 
Thus, our discussions and conclusions are not affected by this imaging issue.

\section{Results}
\subsection{Single-dish maps}\label{sdresult}

\begin{figure*}
\centering
\includegraphics[width=17.5cm]{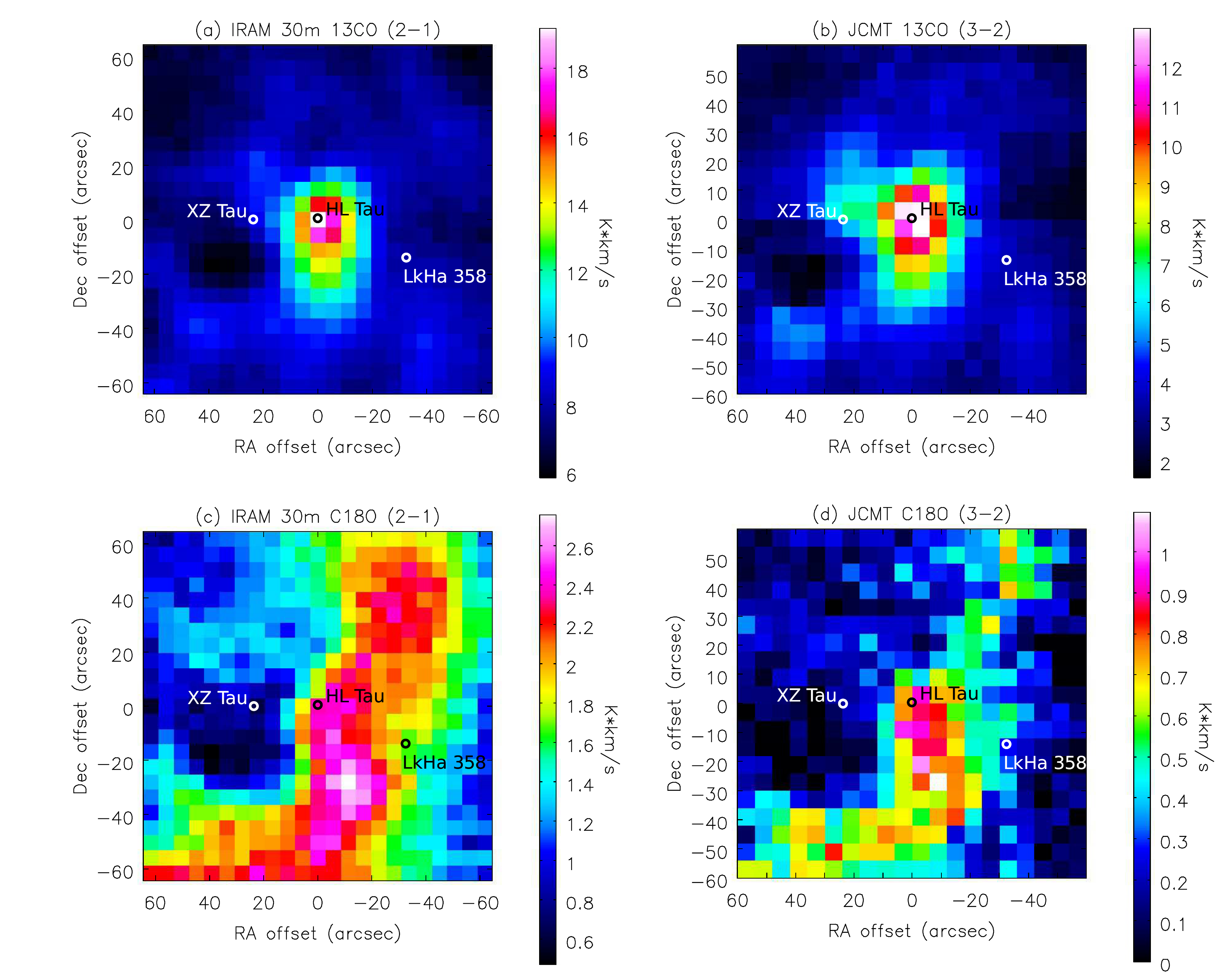}
\caption{Moment 0 maps of the $^{13}$CO (2--1) and C$^{18}$O (2--1) emission (panel a \& c) obtained with the IRAM 30m telescope and  the $^{13}$CO (3--2) and C$^{18}$O (3--2) emission (panel b \& d) obtained with JCMT.}\label{sdmap}
\end{figure*}

\begin{figure*}
\centering
\includegraphics[width=17.5cm]{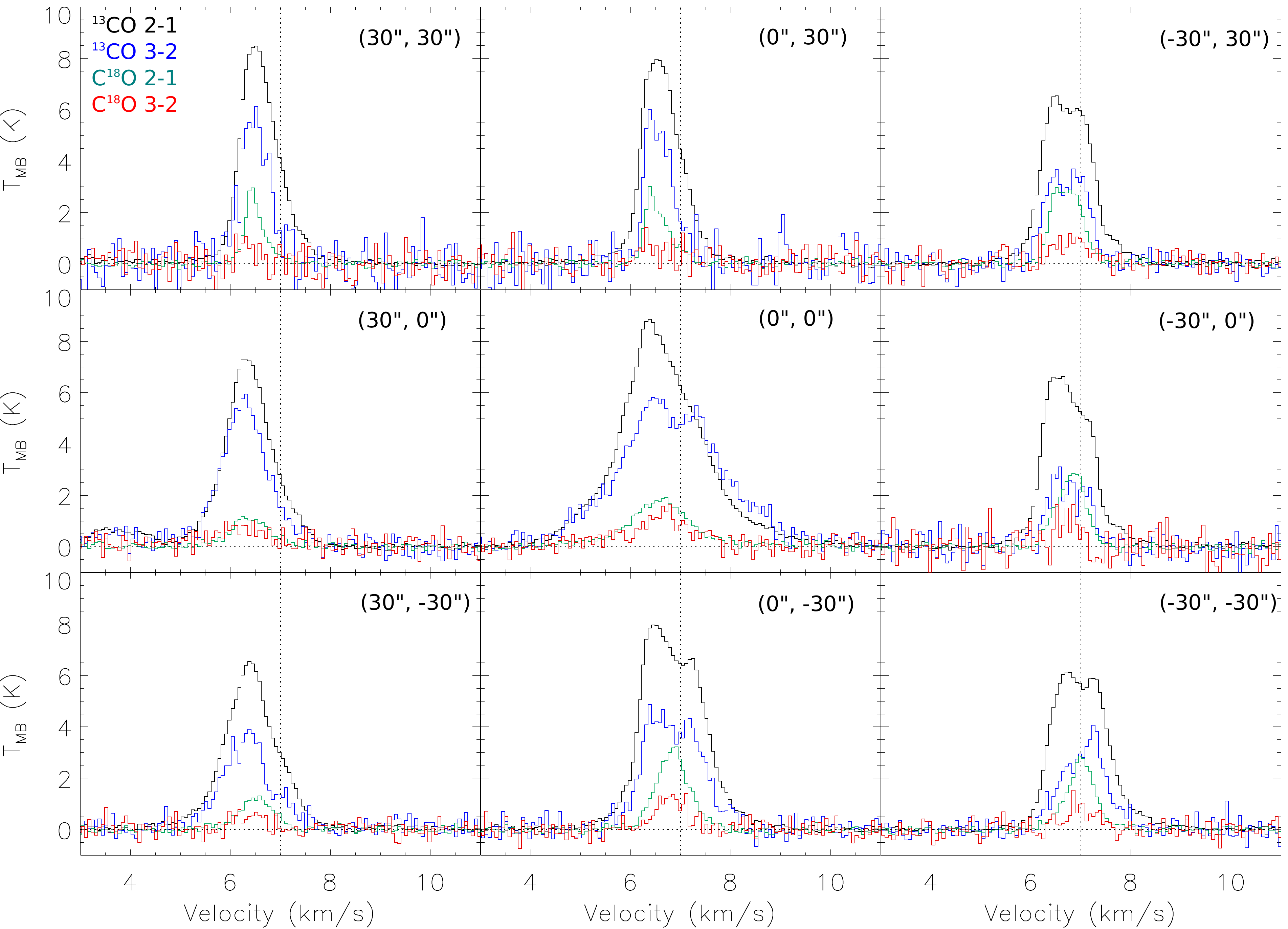}
\caption{Spectra of the $^{13}$CO (2--1; black histograms) and (3--2; blue histograms) emission and the C$^{18}$O (2--1; green histograms) and (3--2; red histograms) emission at nine different positions in the \object{HL~Tau} region obtained with our single-dish observations. The position, where the spectra were extracted, is labeled at the upper right corner in each panel in relative offset with respect to the position of \object{HL~Tau}. Vertical dashed lines denote the systemic velocity of \object{HL~Tau}, $V_{\rm LSR} = 7$ km s$^{-1}$.}\label{specmap}
\end{figure*}

Figure \ref{sdmap} presents the total-integrated intensity (moment 0) maps of the $^{13}$CO and C$^{18}$O emission lines in \object{HL~Tau}.
The intensity distributions of the $^{13}$CO lines are centrally peaked at the position of \object{HL~Tau}. 
In contrast, the C$^{18}$O lines show a filamentary structure from the northwest to the south of \object{HL~Tau}, 
and there is no clear intensity enhancement at the position of \object{HL~Tau}. 
Two T Tauri stars, \object{XZ~Tau} and \object{LkH$\alpha$~358}, are located outside the regions exhibiting the intense $^{13}$CO and C$^{18}$O emission, 
suggesting that they are not associated with dense molecular gas on a scale of thousands of au. 
 
Figure \ref{specmap} presents the $^{13}$CO and C$^{18}$O spectra at nine different positions. 
To compare the spectra of the different emission lines, 
all the data have been first convolved with the same angular resolution of 15\farcs3 and re-gridded to have the same pixel size and channel width. 
The systemic velocity of the circumstellar disk around \object{HL~Tau} is measured to be $V_{\rm LSR}$ of 7 km s$^{-1}$ from the Keplerian rotation of the disk\footnote{To measure the systemic velocity, the peak positions in different velocity channels were measured from the position--velocity diagram along the disk major axis of the $^{13}$CO (2--1) and C$^{18}$O (2--1) emission observed with ALMA at 0\farcs8 resolutions, and a radial velocity profile of Keplerian rotation with two free parameters, stellar mass and systemic velocity, was fitted to these data points.} \citep{Yen17}, 
and we adopt this velocity as the systemic velocity of \object{HL~Tau}. 
The velocity of the intensity peak ($V_{\rm peak}$) of the $^{13}$CO emission at the position of \object{HL~Tau} is 6.4 km s$^{-1}$, which is offset from the systemic velocity of \object{HL~Tau} (Fig.~\ref{specmap}). 
As shown in these spectra, the majority of the emission observed with the single dishes, tracing the large-scale gas and the envelope around \object{HL~Tau}, is at the velocities more blueshifted with respect to the systemic velocity of \object{HL~Tau}.

The intensity ratio of $^{13}$CO (3--2) to (2--1) at the position of \object{HL~Tau} clearly increases as the relative velocity with respect to $V_{\rm peak}$ increases. 
The $^{13}$CO (3--2) to (2--1) intensity ratio is $\sim$0.7 at $V_{\rm peak}$,  
and the intensity ratio becomes close to or larger than unity at the velocities of $V_{\rm LSR} \lesssim 5.6$ km s$^{-1}$ and $V_{\rm LSR} \gtrsim7.2$ km s$^{-1}$.
Similar trends are also seen in the spectra at the other positions, for example, at the blueshifted velocity at (30\arcsec, 30\arcsec) and (30\arcsec, 0\arcsec) and at the redshifted velocity at ($-30$\arcsec, $-30$\arcsec). 
In addition, these spectra show that the C$^{18}$O line wing emission at these higher velocities with respect to $V_{\rm peak}$, where the $^{13}$CO (3--2) to (2--1) intensity ratio is higher, is below the detection level of our observations. 
This change in the intensity ratio of the $^{13}$CO lines suggests that the physical conditions at $V_{\rm peak}$ and those traced by the line wing emission are different. 

On the assumption of the local thermal equilibrium (LTE) condition, the excitation temperature $T_{\rm ex}$ and optical depths ($\tau$) of the $^{13}$CO (3--2) and (2--1) lines can be estimated from
\begin{equation}\label{rteq}
\frac{T_1}{T_2} = (\frac{B_{\nu_1}(T_{\rm ex}) - B_{\nu_1}(T_{\rm bg})}{B_{\nu_2}(T_{\rm ex}) - B_{\nu_2}(T_{\rm bg})})(\frac{1-e^{-\tau_1}}{1-e^{-\tau_2}}), 
\end{equation}
where $B_\nu(T)$ is the Planck function at the frequency $\nu$ and the temperature $T$, and $T_{\rm bg}$ is the cosmic background temperature of 2.73 K. 
Here, $T_1$, $\nu_1$, and $\tau_1$ are the brightness temperature, frequency, and optical depth of the $^{13}$CO (2--1) line, 
and $T_2$, $\nu_2$, and $\tau_2$ are those of the $^{13}$CO (3--2) line. 
We assume that the $^{13}$CO (2--1) and (3--2) lines have the same $T_{\rm ex}$. 
With a given $T_{\rm ex}$, the ratio of $\tau_1$ to $\tau_2$ can be computed following the equations in \citet{Mangum15}. 
Then, the number of unknowns is reduced to two in Eq.~\ref{rteq}, which are $T_{\rm ex}$ and $\tau_1$.
Thus, $T_{\rm ex}$, $\tau_1$, and $\tau_2$ can be derived with Eq.~\ref{rteq} from the brightness temperature of the $^{13}$CO (2--1) and (3--2) lines.

At the position of \object{HL~Tau}, the brightness temperatures of the $^{13}$CO (2--1) and (3--2) lines are both 2.4 K at $V_{\rm LSR}$ of 5.4 km s$^{-1}$ and 5 K at $V_{\rm LSR}$ of 7.3 km s$^{-1}$.
With Eq.~\ref{rteq}, 
$T_{\rm ex}$ is estimated to be 25$\pm$9 K with both $\tau$ of 0.1$\pm$0.1 at $V_{\rm LSR}$ of 5.4 km s$^{-1}$. 
At $V_{\rm LSR}$ of 7.3 km s$^{-1}$, $T_{\rm ex}$ is estimated to be 27$\pm$6 K, and both $\tau$ are 0.3$\pm$0.1.
The uncertainties are estimated from the error propagation of the observational noise.
At these higher velocities with respect to $V_{\rm peak}$, where the intensity ratio is higher, the $^{13}$CO lines are optically thin. 
C$^{18}$O is less abundant than $^{13}$CO with a $^{13}$CO/C$^{18}$O abundance ratio of 5.5--10 in \object{HL~Tau} \citep{Wilson94, Brittain05, Smith15}. 
Thus, the C$^{18}$O lines are expected to be also optically thin at these higher velocities. 
In this case, the ratio of the brightness temperature between the $^{13}$CO and C$^{18}$O lines is approximately their abundance ratio, 
and the expected C$^{18}$O brightness temperature at the higher velocities with respect to $V_{\rm peak}$ is lower than the noise levels in our observations.

As shown in Fig.~\ref{specmap}, the $^{13}$CO (2--1) line is brighter than the $^{13}$CO (3--2) line at the velocity around $V_{\rm peak}$, for example, at $V_{\rm LSR}= 5.6\mbox{--}7.2$ km s$^{-1}$ at the position of \object{HL~Tau}.
The $^{13}$CO (3--2)/(2--1) intensity ratio is expected to decrease with decreasing temperature (Eq.~\ref{rteq}). 
Thus, the gas temperature in the \object{HL~Tau} region is expected to be lower at the velocity around $V_{\rm peak}$.
At $V_{\rm peak}$, which is $V_{\rm LSR}$ of 6.4 km s$^{-1}$, 
the brightness temperatures of the $^{13}$CO (2--1) and (3--2) lines are 8.9 K and 5.7 K, respectively. 
With Eq.~\ref{rteq}, $T_{\rm ex}$ is estimated to be 14$\pm$1 K, and $\tau$ of the $^{13}$CO (2--1) and (3--2) lines are estimated to be 1.5$\pm$0.2 and 1.4$\pm$0.2, respectively.
Here, the signal-to-noise ratios of the emission lines are higher than 10, 
and thus, the dominant source of the uncertainty in the estimate is the absolute flux uncertainties of the IRAM 30m telescope and JCMT, which are 10\%.
Similarly, from the intensity ratio of the C$^{18}$O (3--2) to (2--1) lines, 1.4 K over 1.9 K, at the velocity of the C$^{18}$O intensity peak, $V_{\rm LSR}$ of 6.7 km s$^{-1}$, 
$T_{\rm ex}$ is estimated to be 15$\pm$7 K, and $\tau$ of the C$^{18}$O (2--1) and (3--2) lines are estimated to be 0.3$\pm$0.3 and 0.2$\pm$0.2, respectively.
Therefore, these results show that around \object{HL~Tau} there are relatively warmer gas with a lower column density traced by the line wing emission at the higher velocities with respect to the peak velocity, where the lines are all optically thin, and relatively cooler gas with a higher column density at the velocity close to the peak velocity, where the $^{13}$CO lines are optically thick and the C$^{18}$O lines are optically thin. 

\begin{figure*}
\centering
\includegraphics[width=18cm]{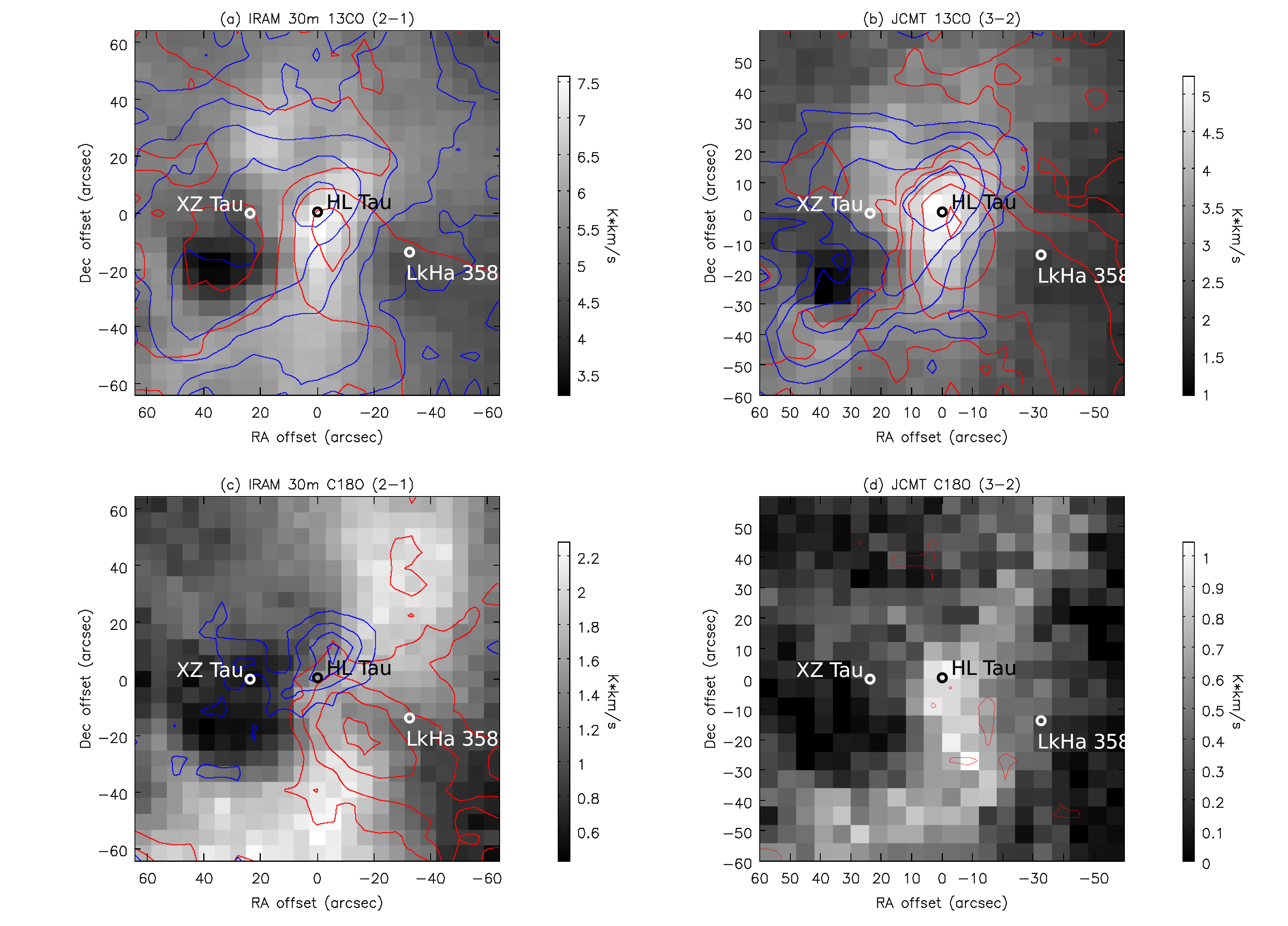}
\caption{Single-dish moment 0 maps of the $^{13}$CO (2--1) and (3--2) emission (panel a \& b) and the C$^{18}$O (2--1) and (3--2) emission (panel c \& d) integrated over different velocity ranges. Grey scales present the integrated intensity in the velocity range close to the velocity of the intensity peak  $V_{\rm LSR} = 6.2\mbox{--}7.1$ km s$^{-1}$. Blue and red contours present the integrated intensity at the higher relative velocities with respect to the peak velocity, $V_{\rm LSR} < 6.2$ km s$^{-1}$ and $V_{\rm LSR} > 7.1$ km s$^{-1}$, respectively. Contours are from 3$\sigma$ and increase in steps of a factor of two. 1$\sigma$ levels of the integrated blue- and redshifted emission are 0.09 and 0.11 K km s$^{-1}$ in (a), are 0.11 and 0.1 K km s$^{-1}$ in (b), and are 0.04 and 0.07 K km s$^{-1}$ in (c). In the C$^{18}$O (3--2) line, there is no clear emission detected at the blueshifted high velocity, and the 1$\sigma$ level of the integrated redshifted emission is 0.1 K km s$^{-1}$.}\label{mom0br}
\end{figure*}

Figure \ref{mom0br} presents the moment 0 maps integrated over different velocity ranges to show the gas distributions at the higher and lower velocities with respect to $V_{\rm peak}$. 
Blue and red contours show the emission at the relative velocities of $>$0.4--0.5 km s$^{-1}$ with respect to $V_{\rm peak}$, corresponding to $V_{\rm LSR} < 6.2$ km s$^{-1}$ and $V_{\rm LSR} > 7.1$ km s$^{-1}$, respectively, 
and the grey scales show the emission at the velocity close to $V_{\rm peak}$, $V_{\rm LSR} = 6.2\mbox{--}7.1$ km s$^{-1}$. 
The C$^{18}$O emission at the velocity close to $V_{\rm peak}$ is elongated from the northwest to the south, identical to the total-integrated intensity (Fig.~\ref{sdmap}). 
Similar north--south elongation can also be seen in the $^{13}$CO emission at the lower velocities. 
At the higher relative velocities with respect to $V_{\rm peak}$, the $^{13}$CO (3--2) and C$^{18}$O (2--1) lines show that the blueshifted emission is located in the northeast and extends toward the east, and that the redshifted emission is located in the southwest and extends toward the southwest. 
A velocity gradient along the same direction, from the northeast to the southwest, is also observed in the $^{13}$CO (2--1) emission at these velocities. 
The C$^{18}$O (3--2) emission is not clearly detected at $V_{\rm LSR} < 6.2$ km s$^{-1}$ and $V_{\rm LSR} > 7.1$ km s$^{-1}$. 
The direction of the velocity gradient observed in the $^{13}$CO (3--2) and (2--1) and C$^{18}$O (2--1) emission at the higher velocities are consistent with that of the bipolar molecular outflow associated with \object{HL~Tau} \citep{Lumbreras14, ALMA15, Klaassen16}. 
These results suggest that the emission lines at $V_{\rm LSR} < 6.2$ km s$^{-1}$ and $V_{\rm LSR} > 7.1$ km s$^{-1}$ on a large scale of $\sim$5000 au observed with the single dishes likely have the contribution from the outflow,  
and that the C$^{18}$O lines are less contaminated by the outflow and better trace the density distribution of the ambient cooler gas with a higher column density around \object{HL~Tau}, compared to the $^{13}$CO lines. 
In addition, the northwest--south elongation of the C$^{18}$O emission is most likely associated with the western edge of the 0.1 pc shell-like structure observed in the $^{13}$CO (1--0) emission with the combined data of the NRAO 12 m telescope and the BIMA array \citep{Welch00}.

\subsection{Combined maps}

\begin{figure*}
\centering
\includegraphics[width=18cm]{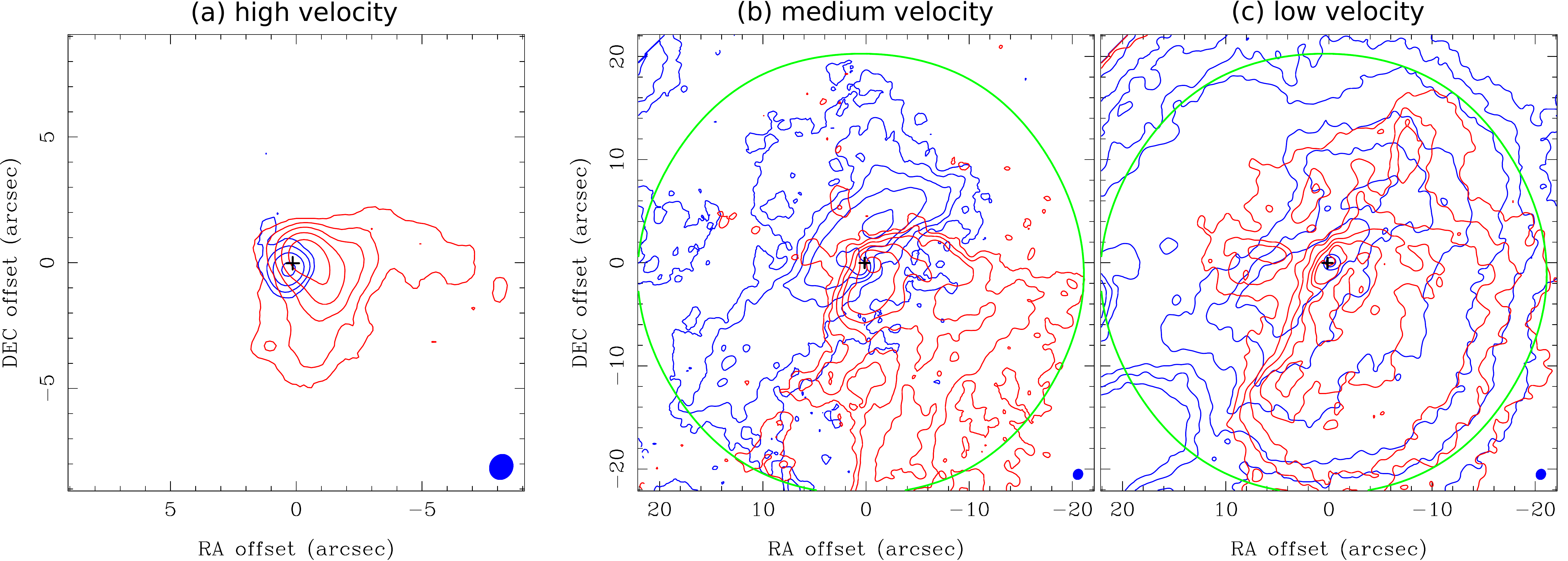}
\caption{Moment 0 maps of the $^{13}$CO (2--1) emission integrated over the different velocity ranges from the combined data obtained with the IRAM 30m, ACA, and ALMA observations. The integrated velocity ranges are high velocities of $V_{\rm LSR}$ = 3--4.7 and 8.5--12 km s$^{-1}$, medium velocities of $V_{\rm LSR}$ = 4.7--5.4 and 7.8--8.5 km s$^{-1}$, and low velocities of $V_{\rm LSR}$ = 5.4--7.1 and 7.1--7.8 km s$^{-1}$. Blue and red contours show the blue- and redshifted emission, respectively. Contour levels are from 5$\sigma$ and increase in steps of a factor of two in (a) and (b), and those are 20$\sigma$, 30$\sigma$, 40$\sigma$, and then in steps of 20$\sigma$ in (c). 1$\sigma$ levels of the integrated blue- and redshifted emission are 6 and 7.4 mJy beam$^{-1}$ km s$^{-1}$ in (a), are 4.6 and 3.4 mJy beam$^{-1}$ km s$^{-1}$ in (b), and are 3.8 and 3.1 mJy beam$^{-1}$ km s$^{-1}$ in (c). Crosses denote the position of \object{HL~Tau}. Filled blue ellipses present the size of the synthesized beam. Green circles show the region where the primary beam response is 0.5.}\label{13comom0}
\end{figure*}

\begin{figure*}
\centering
\includegraphics[width=18cm]{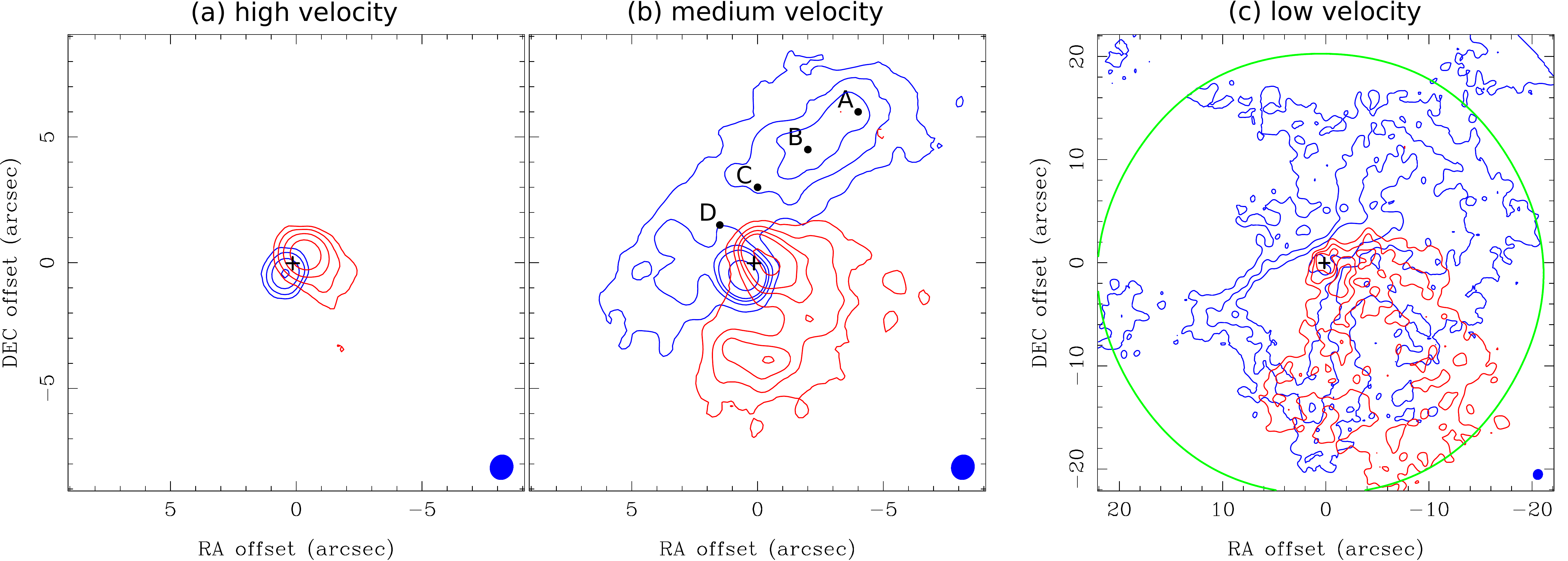}
\caption{Same as Fig.~\ref{13comom0} but for the C$^{18}$O (2--1) emission. 
The integrated velocity ranges are high velocities of $V_{\rm LSR}$ = 1--4 and 8.6--13.4 km s$^{-1}$, medium velocities of $V_{\rm LSR}$ = 4--5.6 and 7.8--8.6 km s$^{-1}$, and low velocities of $V_{\rm LSR}$ = 5.6--7 and 7--7.8 km s$^{-1}$. 
Contour levels are from 5$\sigma$ and increase in steps of a factor two in (a) and (b), and those are 10$\sigma$, 15$\sigma$, 20$\sigma$, 40$\sigma$, and 80$\sigma$ in (c). 1$\sigma$ levels of the integrated blue- and redshifted emission are 4.2 and 5.9 mJy beam$^{-1}$ km s$^{-1}$ in (a), are both 3.2 mJy beam$^{-1}$ km s$^{-1}$ in (b), and are 3.7 and 2.6 mJy beam$^{-1}$ km s$^{-1}$ in (c). 
The positions to extract spectra, which are shown in Fig.~\ref{comspec}, are labeled as A--D in (b).}\label{c18omom0}
\end{figure*}

Figure \ref{13comom0} and \ref{c18omom0} present the moment 0 maps of the $^{13}$CO (2--1) and C$^{18}$O (2--1) emission integrated over different velocity ranges in \object{HL~Tau} obtained by combining the IRAM 30m, ACA, and ALMA data. 
The integrated velocity ranges for the $^{13}$CO line are high velocities of $V_{\rm LSR}$ = 1--4 and 8.6--13.4 km s$^{-1}$, medium velocities of $V_{\rm LSR}$ = 4--5.6 and 7.8--8.6 km s$^{-1}$, and low velocities of $V_{\rm LSR}$ = 5.6--7 and 7--7.8 km s$^{-1}$. 
Those for the  C$^{18}$O line are high velocities of $V_{\rm LSR}$ = 3--4.7 and 8.5--12 km s$^{-1}$, medium velocities of $V_{\rm LSR}$ = 4.7--5.4 and 7.8--8.5 km s$^{-1}$, and low velocities of $V_{\rm LSR}$ = 5.4--7.1 and 7.1--7.8 km s$^{-1}$.
These combined maps do not suffer from the effect of the missing flux, different from the ALMA maps of the $^{13}$CO (2--1) and C$^{18}$O (2--1) emission in \citet{Yen17}.

As discussed in \citet{Yen17}, 
the high-velocity $^{13}$CO and C$^{18}$O components primarily trace the Keplerian rotation of the disk with a radius of $\sim$100 au around \object{HL~Tau}. 
After adding the short-spacing data, 
a possible contamination from the outflow is seen in the redshifted high-velocity $^{13}$CO emission, showing a fan-shape structure extending toward the southwest. 
Such an outflow contamination is not seen in the C$^{18}$O emission. 
At the medium velocities, two arc-like structures are observed in both the $^{13}$CO and C$^{18}$O emission. 
One arc-like structure is blueshifted, and the other is redshifted. 
The blueshifted arc-like structure has a length of 2000 au and stretches from the east to the northwest. 
The redshifted arc-like structure has a length of 1000 au and stretches from the west to the southeast. 
In addition, 
in our combined maps, the extended emission attached to the arc-like structures is also observed in the $^{13}$CO emission, 
and the redshifted arc-like structure becomes less evident, compared to the ALMA map in \citet{Yen17}.
At the low velocities, the $^{13}$CO emission is detected over the entire field of the view. 
For clarity, here in Fig.~\ref{13comom0} and \ref{c18omom0} only contours of the integrated intensity above 20$\sigma$ are plotted. 
At the medium and low velocities, the $^{13}$CO emission exhibit a clear velocity gradient along the northeast--southwest direction, where the northeastern part is blueshifted and the southwestern part is redshifted. 
A similar velocity gradient is also observed in the C$^{18}$O emission at the low velocities. 
The redshifted C$^{18}$O emission at the low velocities additionally shows a fan-shape structure extending toward the southwest with its apex located at the protostellar position. 
This morphology is similar to the outflow in \object{HL~Tau} observed in the CO (1--0) emission with ALMA \citep{Klaassen16} and in the CO (3--2) emission with SMA \citep{Lumbreras14}. 
Therefore, at the low velocities, there is a possible contamination from the outflow.

\begin{figure*}
\centering
\includegraphics[width=17cm]{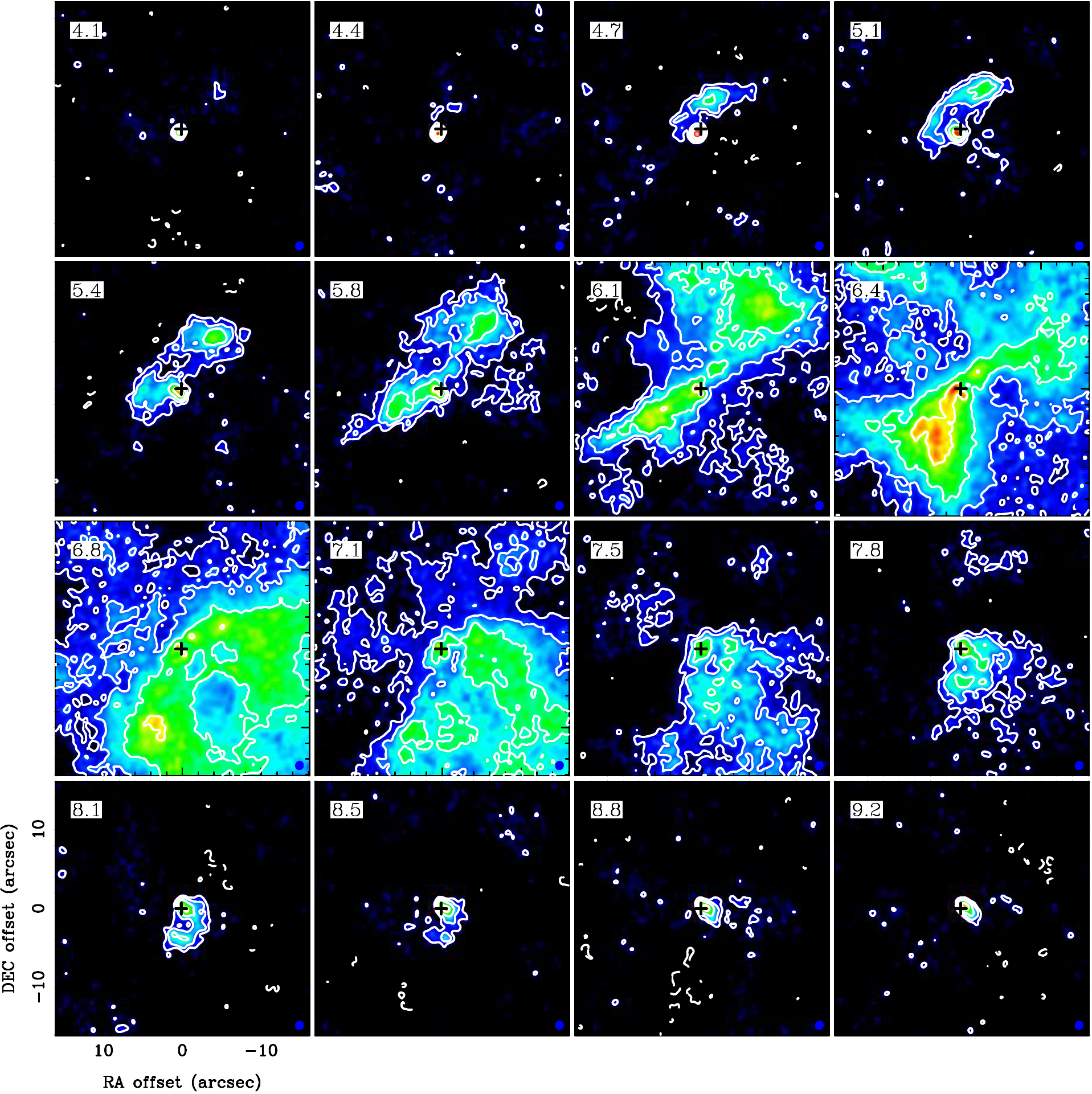}
\caption{Velocity channel maps of the C$^{18}$O (2--1) emission at the medium and low velocities generated from the combined data obtained with the IRAM 30m, ACA, and ALMA observations. Contour levels are from 3$\sigma$ and increase in steps of a factor of two, where 1$\sigma$ is 5.5 mJy Beam$^{-1}$. The central velocity of each channel is labeled at the upper left corner in each panel in units of km s$^{-1}$. Crosses denote the position of \object{HL~Tau}. Blue filled ellipses present the size of the synthetized beam.}\label{c18ochan}
\end{figure*}

With the combined data, 
the arc-like structures, which were only detected in the $^{13}$CO emission at the medium velocities in \citet{Yen17}, are now also detected in the C$^{18}$O emission at similar velocities. 
The entire velocity channel maps in the velocity ranges of the medium and low velocities are shown in Fig.~\ref{c18ochan}.  
As shown in Fig.~\ref{c18ochan}, the arc-like structures are more compact and located closer to the protostar at a higher velocity. 
As the relative velocities with respect to the systemic velocity of \object{HL~Tau} decrease, the extensions of the arc-like structures increase, 
and the arc-like structures gradually merge with the extended emission at the low velocities of $V_{\rm LSR} = 6.1$ and 7.5 km s$^{-1}$. 

Two possible origins of the arc-like structures, infalling material from the extended envelope around \object{HL~Tau} and mass ejection from the \object{HL~Tau} disk, were discussed in \citet{Yen17}.
Gaseous clumps ejected from a disk tend to remain compact as they move away from the disk, as shown in numerical simulations \citep[e.g.,][]{Vorobyov16}.
Thus, the gradual change in the spatial structures from the small to large scales seen in our combined maps suggests that the arc-like structures are unlikely mass ejection from the disk.
In addition, the arc-like structures show a higher velocity at a smaller radius. 
Such a radial velocity profile is opposite to that in structures swept up by outflows, which tends to show a higher velocity at a larger radius \citep{Shu91, Shu00, Lee00}.
The CO outflow in \object{HL~Tau} also shows a higher velocity at a larger radius as observed with SMA \citep{Lumbreras14}.
Therefore, the arc-like structures more likely originate from the extended envelope around \object{HL~Tau}.

\begin{figure}
\centering
\includegraphics[width=9cm]{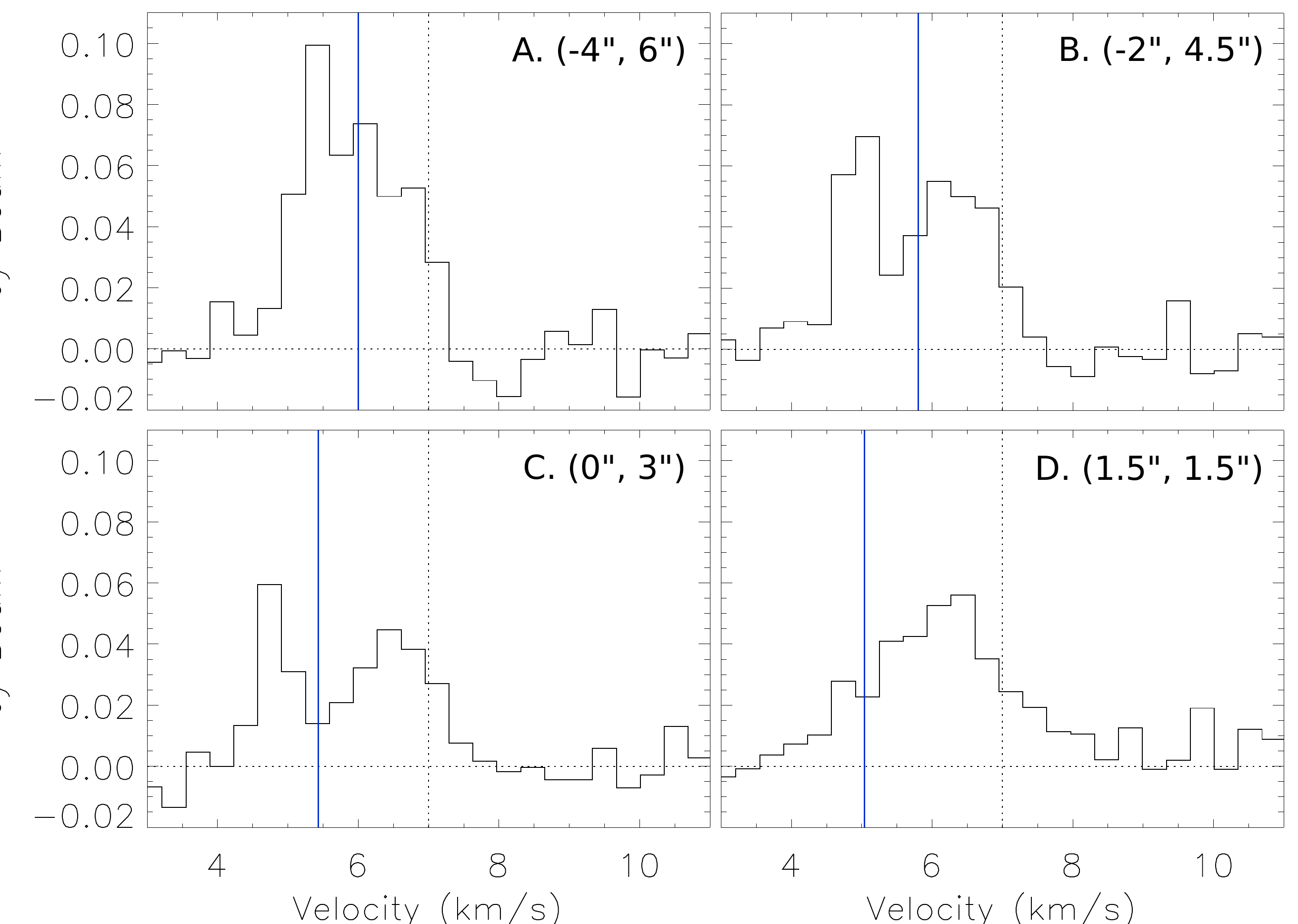}
\caption{Spectra of the C$^{18}$O (2--1) emission extracted from the combined data. The relative offsets of the positions A--D, where the spectra were extracted, with respect to the position of \object{HL~Tau} are shown at the upper right corners in the panels. These positions are also labeled in Fig.~\ref{c18omom0}. Dotted vertical lines denote the systemic velocity of \object{HL~Tau}, $V_{\rm LSR} = 7$ km s$^{-1}$, measured from the Keplerian rotation of its disk \citep{Yen17}. Blue vertical lines show the expected line-of-sight velocities from the model of a free-falling and rotating envelope at those positions \citep{Yen17}.}\label{comspec}
\end{figure}

As discussed in \citet{Yen17}, 
the blueshifted arc-like structure exhibits relative velocities with respect to the systemic velocity of \object{HL~Tau} higher than the expected free-fall velocities$\footnotemark$, 
and there is a possibility that such a velocity excess is caused by the missing flux in the ALMA data and the absorption by the large-scale dense cloud or outflow in the $^{13}$CO emission, which is optically thick at the low velocities. 
The combined map of the C$^{18}$O emission does not suffer from these effects because there is no missing flux and the C$^{18}$O emission is optically thin (Section \ref{sdresult}).
Figure \ref{comspec} presents spectra of the C$^{18}$O emission extracted from the combined data at four different positions along the blueshifted arc-like structure. 
The positions to extract spectra are labeled as A--D in Fig.~\ref{c18omom0}b. 
At the positions A--C, 
there are clearly two velocity components, one peaked at $V_{\rm LSR}$ of $\sim$6.5 km s$^{-1}$ and the other peaked at higher velocities of $V_{\rm LSR} = 4.5\mbox{--}5$ km s$^{-1}$.
The component at the lower velocity is most likely associated with the large-scale ambient gas, 
and the one at the higher velocities is the blueshifted arc-like structure. 
In Fig.~\ref{comspec}, we also plotted the expected line-of-sight velocity from the best-matched model of a free-falling and rotating envelope in \citet{Yen17} at these positions (blue vertical lines).  
This comparison shows that the relative velocities of the blueshifted arc-like structure are indeed higher than the expected free-fall velocity, 
and the velocity excess is not due to the missing flux or the absorption. 

\footnotetext{The expected free-fall velocity was computed with the central stellar mass of 1.8 $M_\odot$ estimated from the Keplerian rotation observed in the C$^{18}$O and $^{13}$CO emission at a 0\farcs8 resolution and the inclination angle of 47$\degr$ estimated from the orientation of the circumstellar disk observed in the 1 mm continuum emission at a 0\farcs03 resolution.}

\section{Analysis}
\subsection{Kinematics of the dense cloud around \object{HL~Tau}}\label{shell}
\object{HL~Tau} is located at the western edge of the shell-like structure with a size of 2$\arcmin$ $\times$ 1\farcm5 ($\sim$11\,000 au in radius) centered at \object{XZ~Tau} observed in the $^{13}$CO (1--0) emission \citep{Welch00}.
The position--velocity (PV) diagrams cutting through this shell show arc-like velocity profiles \citep{Welch00}, 
which can be explained with an expanding shell \citep{Arce11,Offner15}. 
Thus, based on the intensity distribution and the velocity profiles of the $^{13}$CO (1--0) emission, 
\citet{Welch00} suggested that there is an expanding shell driven by \object{XZ~Tau}, which is a T Tauri star and has launched wide-angle wind and molecular outflows \citep{Krist99, Krist08, Zapata15}.

\begin{figure*}
\centering
\includegraphics[width=18cm]{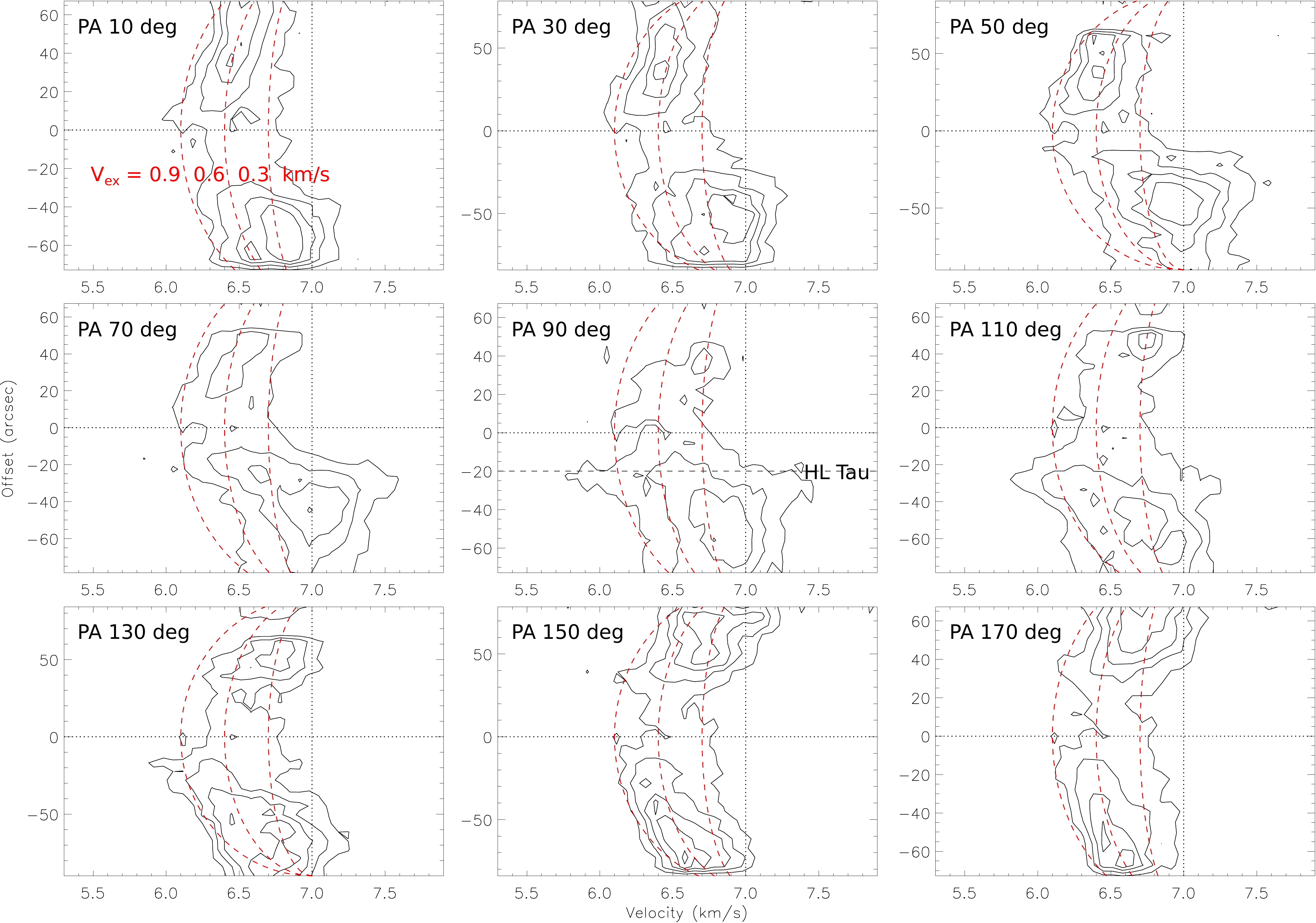}
\caption{Position--velocity diagrams of the C$^{18}$O (2--1) emission obtained with the IRAM 30m observations. The PV diagrams were extracted along different position angles (labeled at the upper left corner in each panel) and passing through the position of \object{XZ~Tau}, which is shown as zero offset and denoted with horizontal dotted lines. Vertical dotted lines present the systemic velocity of \object{HL~Tau} measured from the Keplerian rotation of the \object{HL~Tau} disk. The position of \object{HL~Tau} is at the offset of $-20\arcsec$, denoted as a horizontal dashed line, in the PV diagram along PA of 90\degr. Red dashed curves from left to right in each panel present the expected velocity profiles of the expanding shell with expanding velocities of 0.3, 0.6, and 0.9 km s$^{-1}$, respectively, as labeled in the PV diagram along PA of 10\degr.}\label{sdpv}
\end{figure*}

\begin{figure}
\centering
\includegraphics[width=8.5cm]{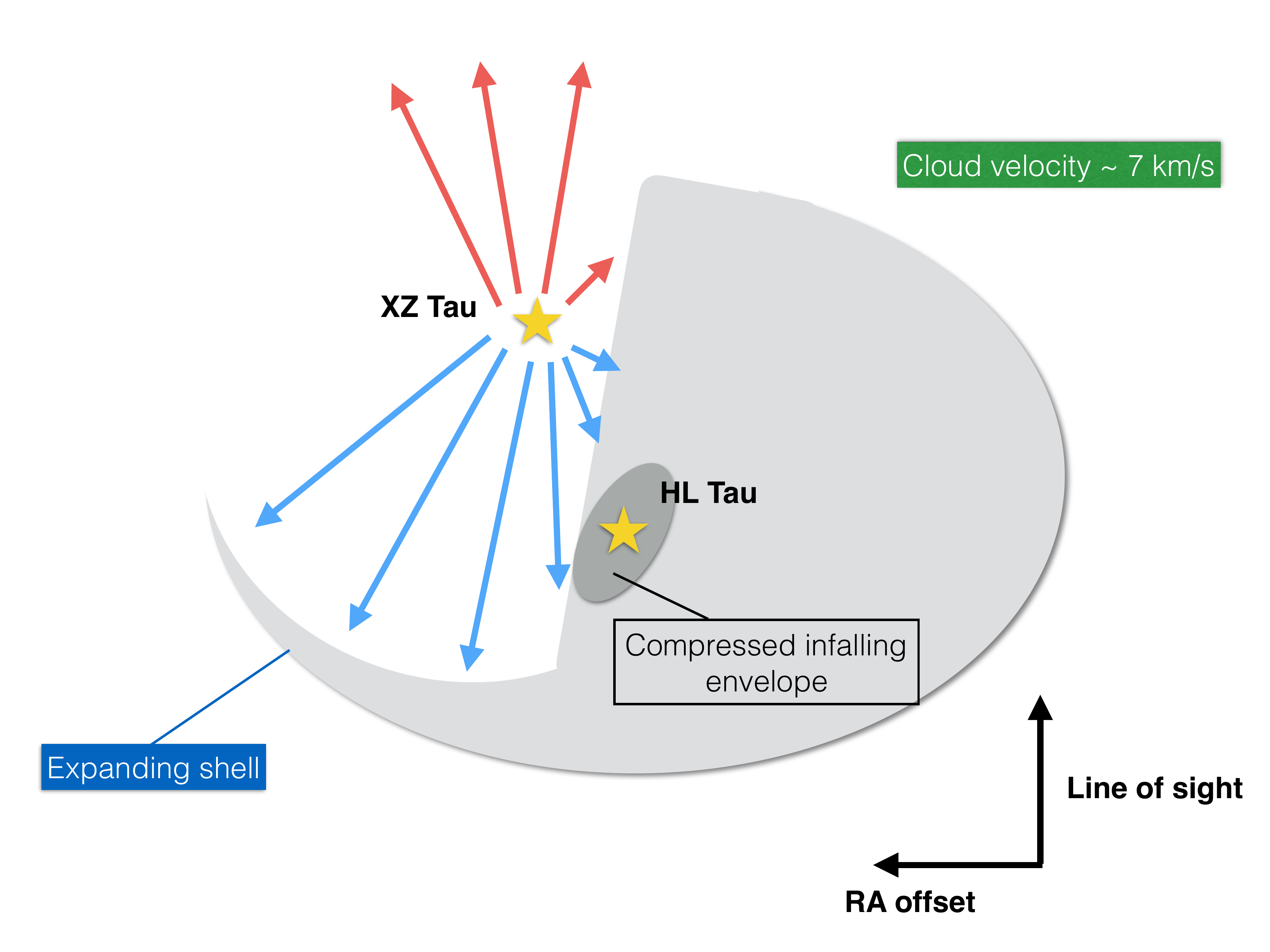}
\caption{Schematic figure of the proposed relative positions between \object{HL~Tau}, \object{XZ~Tau}, and the expanding shell. Horizontal and vertical directions are the directions of RA offset and the line of sight, respectively.}\label{region}
\end{figure}

To examine this scenario of the large-scale expanding shell in the \object{HL~Tau} region with our IRAM 30m data of the C$^{18}$O (2--1) emission, which traces the density distribution of the cold dense gas and has less contamination from the outflow, we extracted a series of PV diagrams centered at the position of \object{XZ~Tau} and along the position angles (PA) from 10$\degr$ to 170$\degr$ in steps of 20$\degr$.
These PV diagrams indeed show arc-like velocity profiles, as expected for an expanding shell (Fig.~\ref{sdpv}). 
We note that there is no redshifted counterpart of the arc-like velocity profiles in these PV diagrams. 
For a spherical expanding shell, we expect to see both blue- and redshifted arc-like velocity profiles in PV diagrams \citep{Arce11}.
This suggests that the shell seen in the $^{13}$CO (1--0) map in \citet{Welch00} is not a spherical shell, 
and that \object{XZ~Tau} is most likely located behind the dense cloud around \object{HL~Tau} and the observed expanding shell.
The dense cloud on the far side, behind \object{XZ~Tau}, is possibly already blowed away by the wind and outflow from \object{XZ~Tau}, 
and thus, no clear redshifted counterpart of the expanding shell is observed.
This was also pointed out by \citet{Welch00}.
In addition, the visual extinction ($A_V$) of \object{XZ~Tau} is measured to be 2.9 magnitude \citep{Furlan06}.
$A_V$ of 2.9 corresponds to a H$_2$ column density of a few $\times$ $10^{21}$ cm$^{-2}$ in the Taurus region on the assumption of the CO abundance of 10$^{-4}$ \citep{Pineda10}, 
suggesting that there is indeed cloud material in front of \object{XZ~Tau}.
In Fig.~\ref{region}, we present a schematic figure of the proposed relative positions between \object{XZ~Tau}, \object{HL~Tau}, and the expanding shell. 

\begin{figure}
\centering
\includegraphics[width=8.5cm]{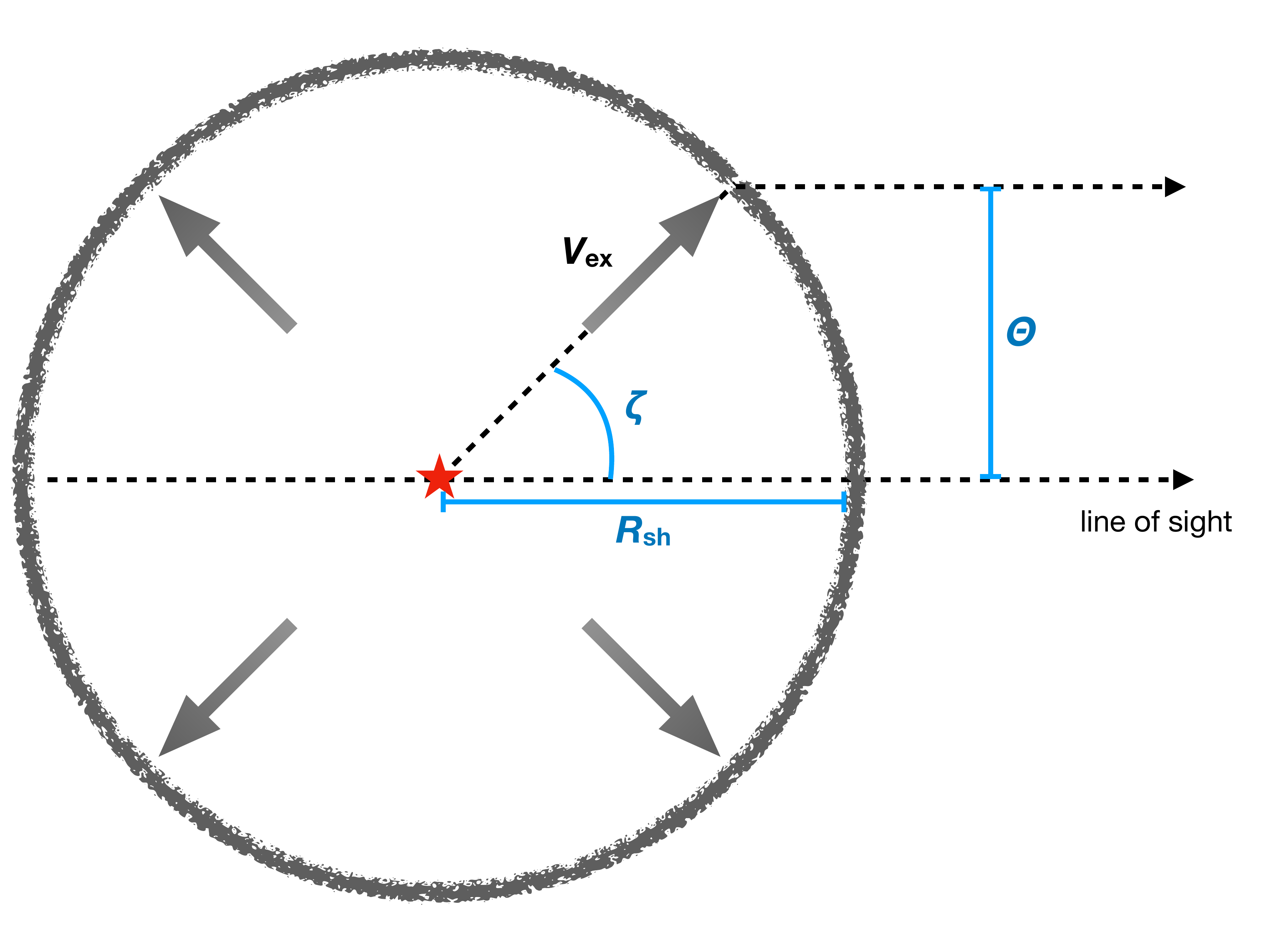}
\caption{Schematic figure of our model of the expanding shell (grey open circle) to illustrate Eq.~\ref{exR} and \ref{exV}. The expansion (grey arrows) is assumed to be spherical and centered at the position of \object{XZ~Tau}, labelled as a red star. $R_{\rm sh}$ is the radius of the expanding shell. $\theta$ is the positional offset with respect to \object{XZ~Tau} on the plane of the sky. $\zeta$ is the angle between the radial direction and the line of sight}\label{expand}
\end{figure}

To measure the velocity of the expanding motion, 
we compared the PV diagrams with the expected velocity profiles for an expanding shell. 
We assume that the observed expanding motion is driven by wind or outflow from \object{XZ~Tau} and is spherical symmetric with respect to the position of \object{XZ~Tau}, 
and that the current radius of the expanding shell ($R_{\rm sh}$) is 90$\arcsec$, which is the radius of the shell observed in the $^{13}$CO emission by \citet{Welch00}. 
In addition, we assume that the systemic velocity ($V_{\rm sys}$) of the large-scale cloud around \object{XZ~Tau} and \object{HL~Tau}, meaning the cloud velocity before being swept up by wind or outflow from \object{XZ~Tau}, is the same as $V_{\rm sys}$ of \object{HL~Tau}, $V_{\rm LSR} = 7$ km s$^{-1}$. 
This $V_{\rm sys}$ is also the centroid velocity of the C$^{18}$O emission observed in the southwestern region, where is further away from \object{XZ~Tau} (Fig.~\ref{specmap}).
Then, the line-of-sight velocity ($V_{\rm los}$) of the expanding motion at a given offset $\theta$ with respect to \object{XZ~Tau} can be computed as, 
\begin{equation}\label{exR}
\zeta = \arcsin \frac{\theta}{R_{\rm sh}},
\end{equation}
and
\begin{equation}\label{exV}
V_{\rm los} = V_{\rm ex}\cos\zeta + V_{\rm sys}, 
\end{equation}
where $\zeta$ is the angle between the radial direction and the line of sight.
A schematic figure of our model of the expanding shell is presented in Fig.~\ref{expand}.

With Eq.~\ref{exR} and \ref{exV}, we computed the expected profiles for the expanding motion at three different velocities, $V_{\rm ex} = 0.3$, 0.6, and 0.9 km s$^{-1}$.
The observed velocity structures can be well described with the expected profiles for the expanding motion, especially in the PV diagrams of PA from 130$\degr$ to 10$\degr$, where the majority of the emission is all within the region enclosed by the expected profiles.
In the PV diagrams of PA from 30$\degr$ to 90$\degr$, the velocity structures at the positive offsets can also be described with these expected profiles, 
while there is an intensity peak at the negative offset and at a velocity close to the assumed cloud velocity, $V_{\rm LSR}$ of 7 km s$^{-1}$, which is offset from the expected profiles for the expanding motion. 
This component is located in the southwestern region in Fig.~\ref{sdmap}, which is brighter and further away from \object{XZ~Tau}. 
Thus, this region could be less affected by the wind or outflow from \object{XZ~Tau} and does not exhibit significant expanding motion.
The position of \object{HL~Tau} is at the offset of $-20\arcsec$ in the PV diagram of PA of 90$\degr$, 
where the majority of the emission is also within the velocity range enclosed by the expected profiles for the expanding motion, 
suggesting that the envelope around \object{HL~Tau} could be also affected by the wind or outflow from \object{XZ~Tau}. 

In summary, the velocity structures observed in the C$^{18}$O emission suggest that the large-scale ambient gas around \object{HL~Tau} is likely expanding at velocities of 0.3--0.9 km s$^{-1}$. 
The estimated $V_{\rm ex}$ is comparable to or lower than that in the $^{13}$CO emission by \citet{Welch00}, 1.2 km s$^{-1}$. 
This difference could be due to different emission lines adopted in the analyses, and our C$^{18}$O (2--1) observations are more sensitive to denser gas compared to the $^{13}$CO observations.

\subsection{Kinematical model of the arc structure in \object{HL~Tau}}
Our combined maps of the C$^{18}$O emission show that the observed velocity in the blueshifted arc-like structure connecting to the disk around \object{HL~Tau} is more blueshifted than the expected line-of-sight velocity from the free-fall motion (Fig.~\ref{comspec}), and this blueshifted velocity excess is not due to the effects of missing flux or optical depth. 
In addition, our single-dish $^{13}$CO and C$^{18}$O spectra at the position of \object{HL~Tau} also shows that the velocities of the intensity peaks are 0.3--0.6 km s$^{-1}$ more blueshifted than the systemic velocity of \object{HL~Tau} (Fig.~\ref{specmap}). 
These blueshifted velocity excesses in the arc-like structure and the protostellar envelope could be naturally explained if the protostellar envelope around \object{HL~Tau} has a relative motion toward us with respect to \object{HL~Tau}.  

To examine this possibility, we constructed a kinematical model of an infalling and rotating protostellar envelope including a relative motion between the envelope and the central protostar. 
Because of this relative motion, 
the infalling and rotational motions in the model envelope cannot be simply described with any conventional models, such as that in \citet[][]{Ulrich76}.
Thus, we constructed a new kinematical model by placing a set of test particles and calculating their trajectories and velocities with time evolution based on the equations of motion.
Then, we projected their positions on the plane of the sky and their velocities on the line of sight to compare with our observations. 

We first assigned each particle an initial velocity computed with the conventional model of an infalling and rotating envelope by \citet{Ulrich76}.
In our model, the mass of the central star is adopted to be 1.8 $M_\sun$, the same as  \object{HL~Tau} \citep{Yen17}, 
and the centrifugal radius is adopted to be 100 au, the observed radius of the disk around \object{HL~Tau} \citep{ALMA15}.
The inclination and position angles of the model envelope are also adopted to be the same as the disk around  \object{HL~Tau}, 47$\degr$ and 138$\degr$, respectively \citep{ALMA15}. 
Then, to introduce a relative motion between the model envelope and the central star, 
we added an additional velocity component ($V_{\rm rel}$) to all the particles in addition to their initial infalling and rotational velocities. 
The direction of this additional velocity is along the line of sight and toward us. 
In summary, in this scenario, 
\object{HL~Tau} first formed out of a protostellar envelope with $V_{\rm sys}$ of 7 km s$^{-1}$, which is the $V_{\rm sys}$ of the disk around \object{HL~Tau}. 
Then after \object{HL~Tau} has accumulated its current stellar mass and the 100 au disk has formed, the protostellar envelope starts to move relatively with respect to \object{HL~Tau}. 
Our kinematical model starts with the protostellar mass of 1.8 $M_\sun$ and the disk radius of 100 au, 
and the relative motion between the envelope and the protostar is added from the start of our model.

We first started with axisymmetric envelope models. 
Test particles were distributed within a radius of 3000 au and 40$\degr$ from the midplane to mimic the flattened envelope as observed in \object{HL~Tau} \citep{Hayashi93}. 
The number density of the test particles in the model envelope is proportional to $r^{-1.5}$, the same as the conventional expectation for an infalling envelope \citep{Shu77}. 
With the initial setup of the particle distribution and the velocity field, 
we then computed the evolution of the model envelopes having different $V_{\rm rel}$ of 0, 0.3, 0.6, and 0.9 km s$^{-1}$.   
Considering that the envelope mass in \object{HL~Tau} is only 0.03--0.06 $M_\odot$ on a scale of 20$\arcsec$ \citep[2800 au; ][]{Hayashi93,Cabrit96} and is 0.13 $M_\odot$ within a radius of 4200 au \citep{Motte01}, less than 10\% of the stellar mass of \object{HL~Tau}, 
the self gravity of the model envelope can be safely ignored. 
In our calculations, the gravity of the central star is assumed to be the only source of the acceleration in the equations of motion, 
and all the particles are assumed to be collisionless. 
In addition, we assume that the particles are accreted onto the disk if they enter the disk region, which is assumed to be a cylinder with a radius of 100 au and a constant height of 10 au.
These accreted particles were removed from the calculations.
We let these model envelopes evolve for $10^4$ yr, 
which is approximately the dynamical time scale for a particle at the outer radius of our model envelope to fall into the center at its free-fall velocity.
As discussed below, this evolutionary time is an arbitrary choice and is not a unique solution to explain the observational results.

After computing the evolution of the model envelopes, we generated synthetic images in the C$^{18}$O emission of our model envelope. 
To compute synthetic image cubes of the C$^{18}$O emission, for a given position, the number density and motional velocity of C$^{18}$O are adopted to be the number density and mean velocity of the test particles at that position in our kinematical models. 
The temperature in our model envelope is adopted from interpolation of the observational estimates. 
The temperature at the disk outer radius of 100 au in \object{HL~Tau} is measured to be 60 K from the CO (1--0) brightness temperature observed with ALMA \citep{Yen16}, on the assumption that the CO (1--0) line is optically thick and its brightness temperature is the kinematic temperature.
On the other hand, the temperature at a radius of 1000 au in the blueshifted arc-like structure in \object{HL~Tau} is estimated to be 15 K from the intensity ratio of the $^{13}$CO (2--1) and C$^{18}$O (2--1) lines \citep{Yen17}, on the assumption of a $^{13}$CO/C$^{18}$O abundance ratio of 10 \citep{Brittain05, Smith15}.
We assume that the radial profile of the temperature in the envelope is a power-law function. 
By interpolating these two temperature measurements at radii of 60 au and 1000 au, 
the temperature profile in our model envelope is adopted to be $T(r) = 15 \times (r/1000{\rm\ au})^{-0.6}\ {\rm K}$.
Then, the C$^{18}$O intensity in the synthetic images was computed with the radiative transfer equation and integrated along the line of sight, 
and we scaled the total number of C$^{18}$O in the model to make the C$^{18}$O intensity in our synthetic images comparable to that in the observations. 
Finally, the synthetic images were convolved with the same synthesized beam as the observations. 
Note that there is no disk component in our synthetic images because our kinematical models do not include a disk.

\begin{figure*}
\centering
\includegraphics[width=16cm]{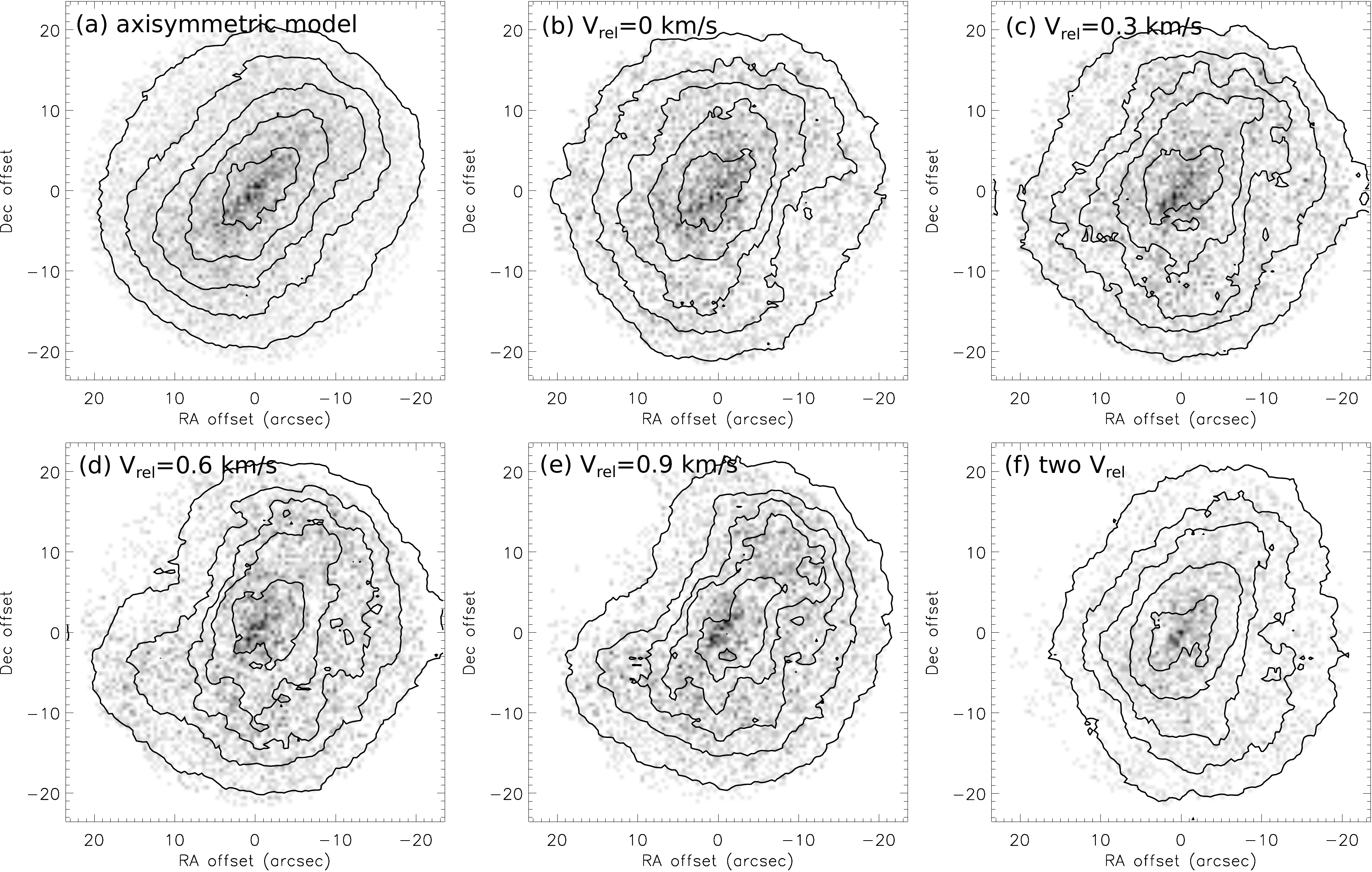}
\caption{Initial particle distributions (grey scale) in our kinematical models projected on the plane of the sky, which represent the maps of the column density of our model envelopes. Panel (a) presents the case of the axisymmetric envelope model. Panel (b)--(e) present the cases of the asymmetric envelope models with different $V_{\rm rel}$ as labeled at the upper left corners. Panel (f) present the model of the asymmetric envelope with the two $V_{\rm rel}$, $V_{\rm rel}$ of 0.9 and 0 km s$^{-1}$ on the far and near sides of the envelope, respectively. Contours show the fraction of the total number of the particles, starting from 10\% in steps of 20\%.}\label{initpar}
\end{figure*}

We note that none of these axisymmetric envelope models forms arc-like structures in their synthetic velocity channel maps of the C$^{18}$O emission.
We found that arc-like structures similar to the observations form in the synthetic velocity channel maps when the initial particle distributions of our kinematical models are not axisymmetric. 
To derive the initial particle distributions in the kinematical models that could result in arc-like structures in the synthetic images, 
we compared the observed velocity channel maps and the evolved particle distributions from our kinematical models projected on the plane of the sky at the same line-of-sight velocities. 
We identified those particles whose positions do not overlap with the observed intensity distribution of the C$^{18}$O emission,
and removed them from the initial particle distributions in our kinematical models. 
An example to derive the initial particle distribution with a given $V_{\rm rel}$ and an evolutionary time is presented in Appendix \ref{modpar}.
The derived initial particle distributions for the kinematical models with the adopted $V_{\rm rel}$ of 0, 0.3, 0.6, and 0.9 km s$^{-1}$ and evolutionary time of 10$^4$ yr projected on the plane of the sky are shown in Fig.~\ref{initpar}.  
Because of the trajectories and velocities of the particle motions in our kinematical models with different $V_{\rm rel}$ are different, 
the derived initial particle distributions for different $V_{\rm rel}$ are different.
In addition, when a different evolutionary time is adopted, the derived initial particle distribution is also different. 
With our model approach, 
an initial particle distribution, which can form arc-like structures after the evolution, can be found for a given set of $V_{\rm rel}$ and an evolutionary time. 
Thus, there is no unique combination of $V_{\rm rel}$ and an evolutionary time to explain the observation.

\begin{figure*}
\centering
\includegraphics[width=17cm]{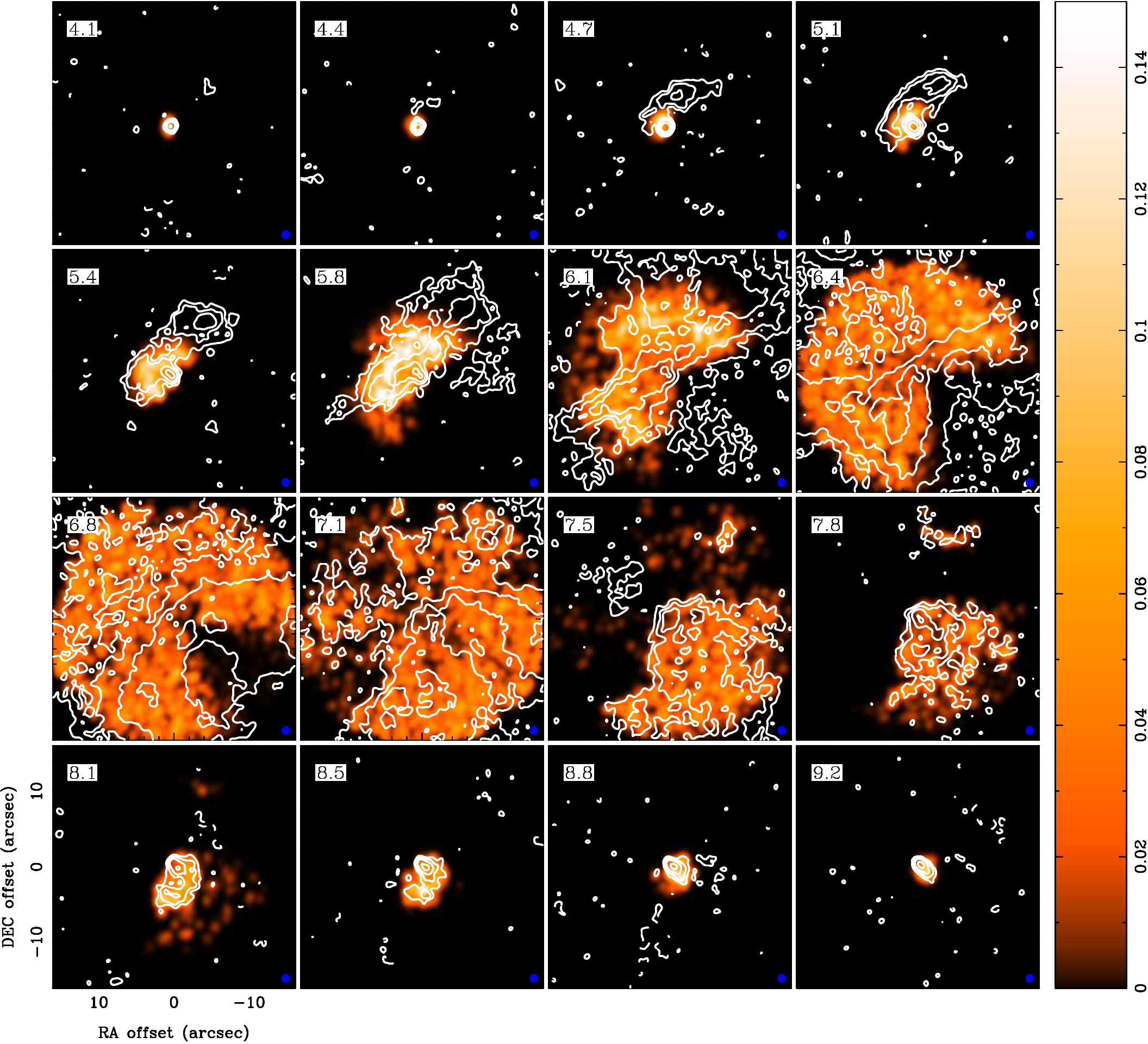}
\caption{Observed (contour) and synthetic (color scale in units of Jy Beam$^{-1}$) velocity channel maps of the C$^{18}$O emission. The synthetic maps were generated from the kinematical model with $V_{\rm rel}$ of 0 km s$^{-1}$ and the asymmetric initial density distribution presented in Fig.~\ref{initpar}. Contour levels are from 3$\sigma$ and increase in steps of a factor of two, where 1$\sigma$ is 5.5 mJy Beam$^{-1}$. The central velocity of each channel is labeled at the upper left corner in each panel in units of km s$^{-1}$. Blue filled ellipses show the size of the synthesized beam.}\label{mod0}
\end{figure*}

\begin{figure*}
\centering
\includegraphics[width=17cm]{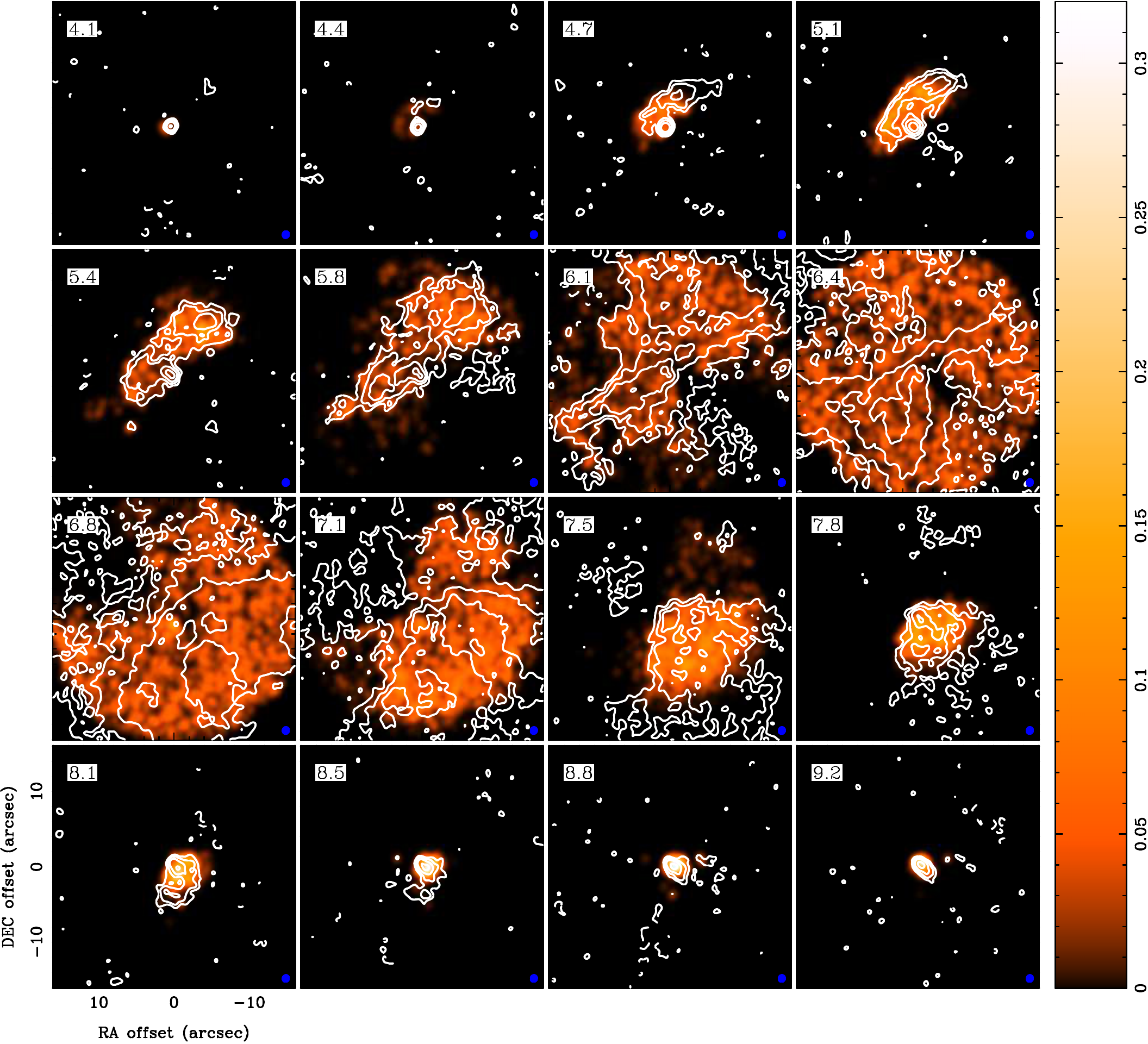}
\caption{Same as Fig.~\ref{mod0} but with $V_{\rm rel}$ of 0.6 km s$^{-1}$.}\label{mod06}
\end{figure*}

Figure~\ref{initpar} shows that when there is less envelope material in the northeastern region (for models with higher $V_{\rm rel}$) and the southwestern region (for models with lower $V_{\rm vel}$), the model envelope could form arc-like structures in its synthetic velocity channel maps after it evolves. 
This distribution with less envelope material in the northeastern and southwestern regions is similar to the observed gas distribution on a scale of thousands of au around \object{HL~Tau}, which shows an elongation from the northwest to the south of \object{HL~Tau}  \citep[Fig.~\ref{sdmap} and ][]{Welch00}.
With the derived initial particle distributions in Fig.~\ref{initpar}, we computed our kinematical models again and generated synthetic images (Fig.~\ref{mod0} and \ref{mod06}).

Our synthetic images of the model with $V_{\rm vel}$ of 0 km s$^{-1}$ can well explain the observed intensity distribution at the redshifted velocity of $V_{\rm LSR}$ of 7.5--9.2 km s$^{-1}$, 
but the blueshifted emission at $V_{\rm LSR}$ of 4.7--5.4 km s$^{-1}$ in the synthetic images is less extended than the observations (Fig.~\ref{mod0}). 
Especially, in the synthetic images with $V_{\rm vel}$ of 0 km s$^{-1}$, there is no counterpart of the northwestern peak in the blueshifted arc-like structure observed at $V_{\rm LSR}$ of 4.7--5.4 km s$^{-1}$. 
In contrast, the synthetic images of the model with $V_{\rm vel}$ of 0.6 km s$^{-1}$ can explain the morphology and velocity of the observed blueshifted arc-like structure at $V_{\rm LSR}$ of 4.7--5.4 km s$^{-1}$, 
but the intensity distribution in the synthetic images at the redshifted velocity of $V_{\rm LSR}$ of 7.5--8.1 km s$^{-1}$ is less extended than the observations (Fig.~\ref{mod06}). 
The synthetic images of the kinematical models with $V_{\rm vel}$ of 0.3 and 0.9 km s$^{-1}$ are shown in Appendix \ref{modimage}. 
The comparison of the observed velocity channel maps with these four models shows that the observed blueshifted emission can be better explained with the models with higher $V_{\rm vel}$ of 0.6 and 0.9 km s$^{-1}$, while the observed redshifted emission with the models with no or lower $V_{\rm vel}$ of 0.3 km s$^{-1}$. 
We also tested other combinations of evolutionary time and initial density distributions. 
We found that our kinematical models with $V_{\rm rel}$ of 0.6 and 0.9 km s$^{-1}$ still form arc-like structures at the blueshifted velocity similar to the observations in the synthetic velocity channel maps, 
and that $V_{\rm rel}$ of 0.6--0.9 km s$^{-1}$ is needed in our kinematical models to explain the high velocity observed in the blueshifted arc-like structure.

\begin{figure*}
\centering
\includegraphics[width=17cm]{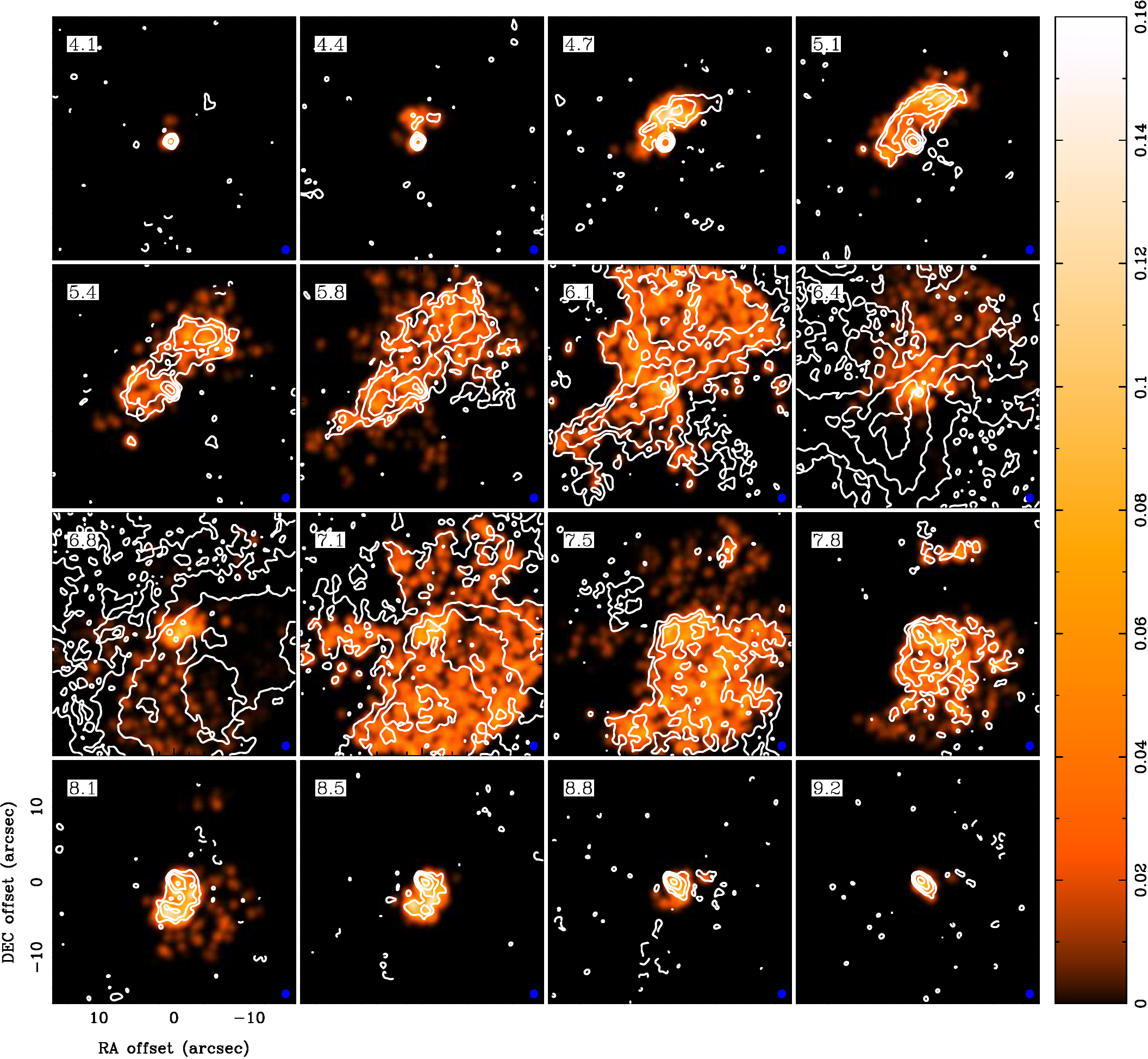}
\caption{Same as Fig.~\ref{mod0} but with the two $V_{\rm rel}$, 0.9 km s$^{-1}$ on the far side and 0 km s$^{-1}$ on the near side of the envelope.}\label{modvmix}
\end{figure*}

In \object{HL~Tau}, the C$^{18}$O emission at blue- and redshifted velocities traces the far and near sides of the infalling protostellar envelope, respectively \citep{Yen17}.
Thus, we constructed an additional model with $V_{\rm rel}$ of 0.9 on the far side and $V_{\rm rel}$ of 0 km s$^{-1}$ on the near side of the envelope, 
and generated synthetic images.
The initial particle distribution of this model with the two $V_{\rm rel}$ is shown in Fig.~\ref{initpar}f, 
and its synthetic images in Fig.~\ref{modvmix}.
Its initial particle distribution is elongated from the northwest to the south, similar to the observed intensity distributions of the C$^{18}$O lines with the single-dish telescopes (Fig.~\ref{sdmap}). 
The synthetic images from this kinematical model can well explain the channel maps at both the blue- and redshifted velocities. 
We note that the model with the two $V_{\rm rel}$ has less emission at $V_{\rm LSR}$ of 6.4 and 6.8 km s$^{-1}$ compared to the observations. 
This could be due to the large-scale ambient gas with the peak velocity of 6.7 km s$^{-1}$ around the protostellar envelope (Fig.~\ref{sdmap} and \ref{specmap}) not included in our model. 

\section{Discussion}
\subsection{Possible relative motion between \object{HL~Tau} and its protostellar envelope}
The observed velocities and morphologies of the arc-like structures connected to the disk around \object{HL~Tau} cannot be explained with any conventional model of an infalling and rotating envelope, and the observed velocity in the blueshifted arc-like structure is higher than the expectation from the free-fall motion.
With our new kinematical models, the high velocity observed in the blueshifted arc-like structure connecting to the disk can be explained by including a relative motion between \object{HL~Tau} and its protostellar envelope in addition to the original infalling and rotational motions. 
The morphologies of the arc-like structures can also be reproduced with our kinematical models with asymmetric initial density distributions (Fig.~\ref{initpar}). 
In addition, the comparison between the observations and our kinematical models with the relative motions at different velocities suggests that the far side of the envelope has a higher additional relative velocity with respect to \object{HL~Tau}, while the near side of the envelope has a lower or no additional relative velocity. 
Our kinematical models show that when the far side of the model envelope has an additional relative velocity of 0.6--0.9 km s$^{-1}$ along the line of sight with respect to the central protostar, 
the observed velocity in the blueshifted arc-like structure can be well reproduced. 

There are two possibilities of the protostellar envelope moving relatively with respect to \object{HL~Tau}, 
the envelope suffered an external impact or \object{HL~Tau} being attracted by other nearby objects.  
Our single-dish results support the presence of an expanding shell driven by \object{XZ~Tau}, as suggested by \citet{Welch00}.  
The expanding velocity is estimated to be 0.3--0.9 km s$^{-1}$ with our C$^{18}$O (2--1) observations and to be 1.2 km s$^{-1}$ with the $^{13}$CO (1--0) observations by \citet{Welch00}. 
We note that the velocity of the wind or outflow driving this large-scale expansion can be much higher than this expanding velocity \citep{Welch00}.
\object{HL~Tau} is located at the edge of this expanding shell,  
and our single-dish $^{13}$CO and C$^{18}$O spectra show that the ambient gas on a 1000 au scale around \object{HL~Tau} is strongly blueshifted with respect to the systemic velocity of \object{HL~Tau} by 0.3--0.6 km s$^{-1}$. 
These results suggest that the protostellar envelope around \object{HL~Tau} can be impacted by the large-scale expanding motion (Section \ref{shell}), 
which was also suggested by \citet{Welch00}.
The estimated expanding velocity is comparable to the required velocity of the relative motion between \object{HL~Tau} and its protostellar envelope to explain the observed velocity structures in the combined maps with our kinematical models. 
In addition, our kinematical models suggest that the relative velocity between \object{HL~Tau} and its protostellar envelope is higher on the far side and lower on the near side of the envelope. 
The candidate driving source of the large-scale expanding motion, \object{XZ~Tau}, is most likely located behind \object{HL~Tau} along the direction of the line of sight (Fig.~\ref{region}).
Thus, the far side of the envelope, where is closer to the driving source, likely gains more momentum from the large-scale expanding motion and has a higher additional relative velocity, while the near side is shielded by the far side and is less impacted by the expanding motion. 
Therefore, the observed velocity structures and morphology of the protostellar envelope are consistent with the expectation from this scenario of the impact by the large-scale expanding motion driven by wind or outflow from the T Tauri star \object{XZ~Tau}.

In this scenario, the three-dimensional distance between \object{HL~Tau} and the driving source of the large-scale expansion, \object{XZ~Tau}, is 90$\arcsec$ (12600 au) based on the observed radius of the expanding shell (Fig.~\ref{expand}).
If the expanding shell originates from the position of \object{XZ~Tau} with a constant expanding velocity of 1 km s$^{-1}$, 
it takes $5 \times 10^4$ yr for the expanding shell to reach the position of \object{HL~Tau}. 
On the other hand, the expansion is likely driven by the wind or outflow from \object{XZ~Tau} whose velocity is much higher than the expanding velocity of the shell. 
With a wind or outflow velocity of 10 km s$^{-1}$, the wind or outflow reaches the position of \object{HL~Tau} $5 \times 10^3$ yr after its launch. 
Depending on the initial density distribution of the ambient cloud around \object{HL~Tau} and \object{XZ~Tau}, 
it is most likely that the wind or outflow from \object{XZ~Tau} first travels at a high velocity of tens of km s$^{-1}$ and impacts the ambient cloud to form the expanding shell, 
and then the shell expands at a low velocity of 0.6--1.2 km s$^{-1}$, reaches and impacts the envelope around \object{HL~Tau}.
Therefore, the impact on the envelope around \object{HL~Tau} by the large-scale expansion possibly occurred between $5 \times 10^3$ and $5 \times 10^4$ ago. 
In our kinematical model, we adopt an evolutionary time of 10$^4$ yr, and it is not a unique time scale to explain the observational results.
Thus, considering the time scales of the propagation of the wind or outflow from \object{XZ~Tau} and the large-scale expansion, 
the scenario of the external impact is indeed possible.

Alternatively, the relative motion between \object{HL~Tau} and its protosteller envelope could be due to \object{HL~Tau} being attracted by nearby objects, such as \object{XZ~Tau}, and moving relatively to its envelope. 
However, this scenario cannot explain why the protostellar envelope is not also being attracted by the same gravitational source and moves together with \object{HL~Tau}. 
The protostellar envelope around \object{HL~Tau} is embedded in the filamentary structure, 
and there is not much gas around \object{XZ~Tau} (Fig.~\ref{sdmap}). 
The pressure gradient of such a gas distribution is unlikely to support the protostellar envelope not to be attracted and move together with \object{HL~Tau}, if there is indeed a gravitational source to drag \object{HL~Tau}. 
Therefore, this scenario of \object{HL~Tau} being attracted and leaving its protostellar envelope is unlikely.

\subsection{Impact on the kinematics of protostellar envelopes by nearby outflow feedback}
Our results hint that the protostellar envelope around \object{HL~Tau} is impacted by the large-scale expanding motion driven by wind or outflow from the nearby young stellar object. 
Such an impact can affect the gas kinematics and thus evolution of the protostellar envelope around \object{HL~Tau}. 
We have computed our kinematical models with different relative velocities between \object{HL~Tau} and its envelope for an evolutionary time longer than 10$^4$ yr. 
In the model with $V_{\rm rel}$ of 0 km s$^{-1}$, 
the entire envelope is eventually accreted onto the disk, meaning that all the test particles enter the disk region,  within a time scale of 2 $\times$ 10$^4$ yr, as expected in the conventional model of an infalling and rotating envelope \citep[e.g.,][]{Ulrich76}.
On the other hand, 
in the model envelope with $V_{\rm rel}$ of 0.3, 0.6, and 0.9 km s$^{-1}$, 
there are 60\%, 80\%, and 90\% of particles, respectively, which do not encounter the disk region and leave the system. 
Because of the additional relative velocity, these particles become gravitationally unbounded, and their centrifugal radii become larger than the disk radius of 100 au.
As a result, they eventually pass by the disk when they infall toward the center. 

Thus, after the protostellar envelope around  \object{HL~Tau} is impacted by the large-scale expanding motion, 
if there is no other mechanism, such as magnetic braking, to transfer the angular momenta of the envelope material away, 
a part of the envelope is unlikely to be accreted onto the disk around \object{HL~Tau}. 
On the other hand, magnetic braking is expected to be inefficient in the later evolutionary stage when protostellar envelopes start to dissipate \citep[e.g.,][]{Machida11}.
In our kinematical models with the two $V_{\rm rel}$, which can well explain the observed velocity channel maps at both the blue- and redshifted velocities, 
only the half of the model envelope is accreted onto the disk. 
Therefore, the mass infalling rate from the envelope onto the disk in \object{HL~Tau} likely decreases by a factor of two due to the impact by the large-scale expanding motion.

Similar impact of outflow launched by nearby protostars on protostellar envelopes have also been observed in the Class I protostar, \object{L1551~NE}.
The protostellar envelope around \object{L1551~NE} shows the signatures of dissipating motion, 
which is likely caused by the impact of the outflow from the nearby protostar, \object{L1551~IRS5} \citep{Takakuwa15}.  
It is expected that there is no further mass infall onto the circumbinary disk around \object{L1551~NE} \citep{Takakuwa15}.
\object{L1448C(S)} is also a candidate protostar having such impact of outflow on its protostellar envelope \citep{Hirano10}. 
In the \object{L1448} region, the position of \object{L1448C(S)} on the plane of the sky overlaps with the outflow launched by \object{L1448C(N)}. 
\object{L1448C(S)} is surrounded by a small amount of circumstellar material ($<$0.01 $M_\sun$) traced by the 860 $\mu$m continuum and less obscured in infrared, compared to \object{L1448C(N)}, suggesting that the envelope around \object{L1448C(S)} can be stripped away by the outflow of \object{L1448C(N)} \citep{Hirano10}. 

Such influence of the outflow feedback on protostellar sources may not be rare. 
Expanding shells similar to that around \object{XZ~Tau} have often been observed in star-forming regions. 
With Nobeyama 45m observations in the CO (1--0) and $^{13}$CO (1--0) emission, 42 expanding shells were identified in the massive star-forming region, Orion A,  
and the radii and expanding velocities of these shells range from 0.05 pc to 0.85 pc and from 0.8 km s$^{-1}$ to 5 km s$^{-1}$, repsectively \citep{Feddersen18}.
Many expanding shells have also been observed in low- or intermediate-mass star-forming regions. 
In the Perseus star-forming region, 
12 expanding shells with radii of 0.1 pc to  3 pc and expanding velocities of 1 to 6 km s$^{-1}$ were identified with the COMPLETE survey in the CO (1--0) and $^{13}$CO (1--0) lines \citep{Arce11}.  
In the Taurus star-forming region, 
37 expanding shells were found in the FCRAO observations in the CO (1--0) and $^{13}$CO (1--0) lines \citep{Li15}.
These expanding shell are suggested to be driven by young stellar objects \citep{Arce11, Li15, Feddersen18}. 

The interaction between protostellar envelopes and associated outflows in protostellar sources has been suggested to affect the structures of the envelopes and to limit the volume of the infalling region, 
and eventually the protostellar envelopes dissipate and the mass infalling rates decrease \citep{Takakuwa03, Arce06, Machida12}. 
Our results suggest an additional path to decrease mass infalling rates in protostellar envelopes by outflow feedback from nearby young stellar objects.

\section{Summary}
We have conducted observations toward \object{HL~Tau} in the $^{13}$CO (3--2) and C$^{18}$O (3--2) lines with JCMT and in the $^{13}$CO (2--1) and C$^{18}$O (2--1) lines with the IRAM 30m telescope and the ACA 7-m array, and generated combined images with the IRAM 30m, ACA, and ALMA data.
With the single-dish and interferometric data, we have studied the gas motions on a 0.1 pc scale in the \object{HL~Tau} region and in the protostellar envelope on a 1000 au scale around \object{HL~Tau}. 
Our main results are summarized below. 
\begin{enumerate}
\item{Our single-dish images of the C$^{18}$O (3--2) and (2--1) emission show that \object{HL~Tau} is located in a large-scale (0.1 pc)  filamentary structure elongated from the northwest to the south of \object{HL~Tau}. 
On the contrary, the $^{13}$CO (3--2) and (2--1) emission is centrally peaked at the position of \object{HL~Tau} and does not exhibit clear elongation. 
The $^{13}$CO intensity distribution most likely has contribution from the outflow associated with \object{HL~Tau}, while the C$^{18}$O emission is less contaminated by the outflow and traces the structures of the large-scale molecular cloud around \object{HL~Tau}.
The comparison of the intensity ratios between (3--2) and (2--1) transitions of the C$^{18}$O  and $^{13}$CO lines shows that the C$^{18}$O emission is optically thin, 
while the $^{13}$CO emission is optically thick at the peak velocity ($V_{\rm LSR} = 6.4$ km s$^{-1}$) and is optically thin at the line wing ($V_{\rm LSR} < 5.6$ km s$^{-1}$ and $> 7.2$ km s$^{-1}$).}
\item{Our combined images generated from the IRAM 30m, ACA, and ALMA data show that the C$^{18}$O (2--1) and $^{13}$CO (2--1) emission lines at the high velocities, $\Delta V \lesssim -3$ km s$^{-1}$ and $\gtrsim 1.5$ km s$^{-1}$ with respect to the systemic velocity of \object{HL~Tau}, primarily trace the rotation of the circumstellar disk on a 100 au scale around \object{HL~Tau}. 
Blue- and redshifted arc-like structures are observed at the medium velocities, $|\Delta V| > 0.8\mbox{--}1.4$ km s$^{-1}$. 
The sizes of the arc-like structures gradually increase with decreasing relative velocity, 
and the arc-like structures emerge with the large-scale molecular cloud at the low velocity, $|\Delta V| < 0.8\mbox{--}1.4$ km s$^{-1}$, 
suggesting that the arc-like structures are most likely formed by infalling material from the ambient gas around \object{HL~Tau}. 
In addition, the C$^{18}$O (2--1) spectra of the blueshifted arc-like structure from the combined data show that its relative velocity with respect to the systemic velocity of \object{HL~Tau} is higher than the expected free-fall velocity, 
and this velocity excess is not due to the effects of missing flux or absorption.}
\item{Position--velocity diagrams of the C$^{18}$O (2--1) emission passing through the position of \object{XZ~Tau} along several different position angles obtained with the IRAM 30m observations show arc-like velocity profiles at the blueshifted velocities, 
suggesting that the large-scale molecular cloud in the \object{HL~Tau} region has expanding motion, possibly driven by the wind or outflow from \object{XZ~Tau}, as suggested by \citet{Welch00}.
The velocity of the expanding motion is estimated to be 0.3--0.9 km s$^{-1}$.
\object{HL~Tau} is located at the edge of this expanding shell.
No signature of the expanding motion was observed at the redshifted velocity, suggesting that \object{XZ~Tau} is located behind \object{HL~Tau} along the line of sight, 
and the gas on the rear side of \object{XZ~Tau} may already be blowed away.  
}
\item{We constructed kinematical models of an infalling and rotating envelope including a relative motion between the central star and the envelope at velocities of 0, 0.3, 0.6, and 0.9 km s$^{-1}$. 
We computed the time evolution of the model envelope and generated synthetic images in the C$^{18}$O (2--1) emission.
We found that when the relative motion is included, our kinematical models can explain the observed high velocity in the blueshifted arc-like structure, 
and the morphologies of the observed arc-like structures can also be well reproduced with our models having an asymmetric initial density distribution.
The comparison between our models and the observations suggests that in addition to the infalling and rotational motions in the protostellar envelope, the far side of the envelope is moving relatively toward us with respect to \object{HL~Tau} at a velocity of 0.6--0.9 km s$^{-1}$, while the near side of the envelope has no or a slower relative motion at a velocity of 0.3 km s$^{-1}$.}
\item{The additional relative velocity in the protostellar envelope with respect to \object{HL~Tau} is comparable to the expanding velocity of the large-scale expanding shell.
In addition, the additional relative velocity on the far side of the envelope, where is closer to the candidate driving source of the large-scale expanding shell, is higher than that on the near side. 
These results suggest that the relative motion between \object{HL~Tau} and its envelope could be caused by the impact of the large-scale expanding motion driven by the wind or outflow from \object{XZ~Tau}.
Such the impact likely affects the kinematics and structures of the protostellar envelope around \object{HL~Tau}, and a part of the envelope likely becomes gravitational unbounded and is unlikely to be accreted onto the disk around \object{HL~Tau}. 
If there is no other mechanism to reduce the excess angular momentum caused by the impact, 
our kinematical models suggest that the mass infalling rate from the envelope onto the disk in \object{HL~Tau} is expected to decrease by a factor of two.
Our results could demonstrate the possibility of decrease in mass infalling rates in protostellar envelopes due to outflow feedback from nearby young stellar objects.}
\end{enumerate}

\begin{acknowledgements} 
This paper makes use of the following ALMA data: ADS/JAO.ALMA\#2015.1.00551.S. ALMA is a partnership of ESO (representing its member states), NSF (USA) and NINS (Japan), together with NRC (Canada), MOST and ASIAA (Taiwan), and KASI (Republic of Korea), in cooperation with the Republic of Chile. The Joint ALMA Observatory is operated by ESO, AUI/NRAO and NAOJ.  
The JCMT data were obtained under program ID M17AP086. 
JCMT is operated by the East Asian Observatory on behalf of The National Astronomical Observatory of Japan; Academia Sinica Institute of Astronomy and Astrophysics; the Korea Astronomy and Space Science Institute; the Operation, Maintenance and Upgrading Fund for Astronomical Telescopes and Facility Instruments, budgeted from the Ministry of Finance (MOF) of China and administrated by the Chinese Academy of Sciences (CAS), as well as the National Key R\&D Program of China (No. 2017YFA0402700). Additional funding support is provided by the Science and Technology Facilities Council of the United Kingdom and participating universities in the United Kingdom and Canada. 
The Starlink software (Currie et al.~2014) is currently supported by the East Asian Observatory.
This work is based on observations carried out under project number 143-16 with the IRAM 30m telescope. IRAM is supported by INSU/CNRS (France), MPG (Germany) and IGN (Spain).
We thank all the JCMT, IRAM 30m, and ALMA staff supporting this work.
This work was supported by NAOJ ALMA Scientific Research Grant Numbers 2017-04A.
S.T. acknowledges a grant from the Ministry of Science and Technology (MOST) of Taiwan (MOST 102-2119- M-001-012-MY3), 
and JSPS KAKENHI Grant Numbers JP16H07086 and JP18K03703 in support of this work.
\end{acknowledgements}

\begin{appendix}
\section{Initial particle distributions in our kinematical models}\label{modpar}
\begin{figure*}
\centering
\includegraphics[width=18cm]{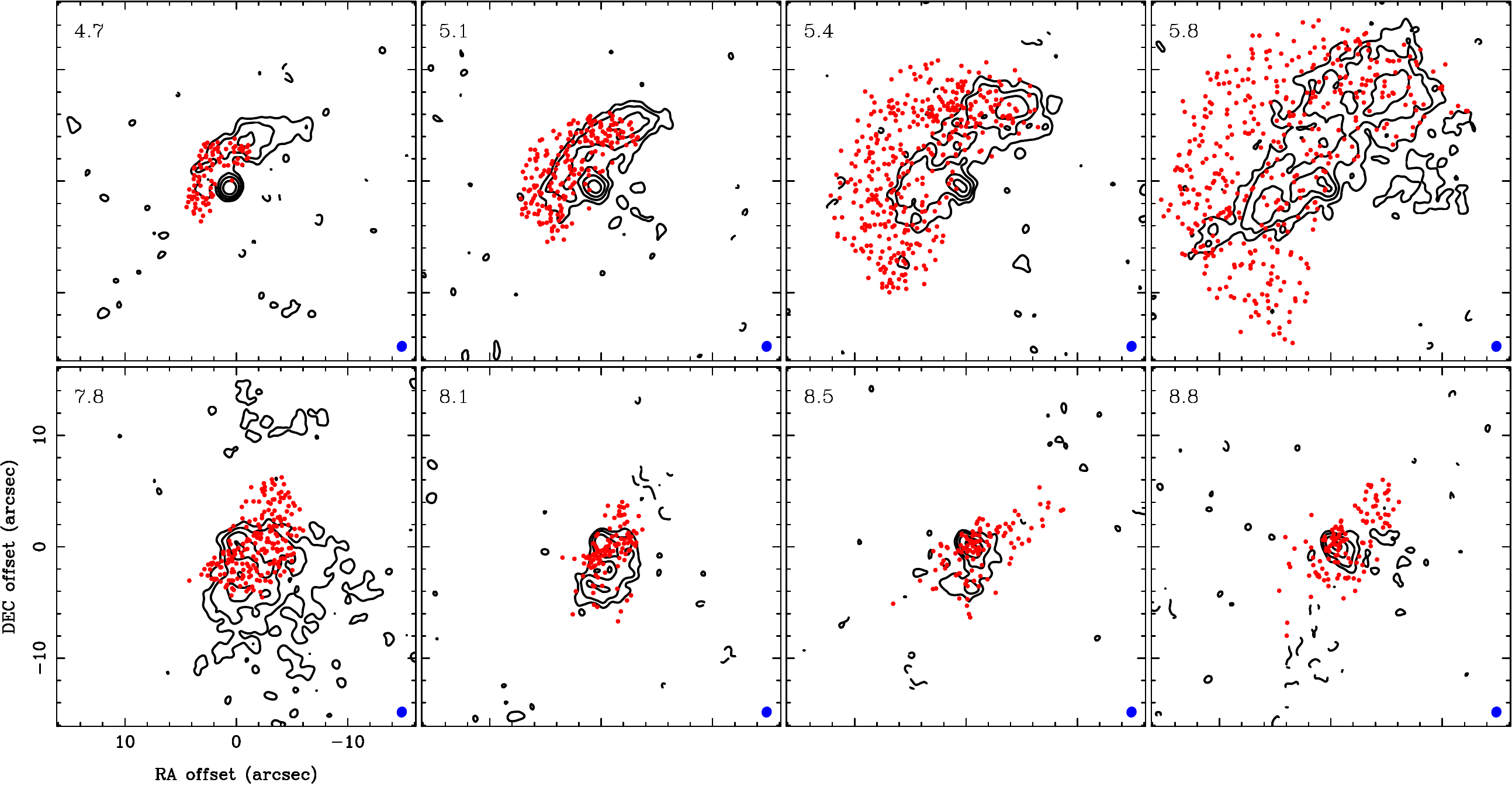}
\caption{Observed velocity channel maps of the C$^{18}$O emission (contour) overlaid with the particle distributions (red dots) at the same line-of-sight velocities from our kinematical model with $V_{\rm rel}$ of 0.6 km s$^{-1}$ after an evolutionary time of 10$^4$ yr. Contour levels are from 3$\sigma$ and increase in steps of a factor of two, where 1$\sigma$ is 5.5 mJy Beam$^{-1}$. The central velocity of each channel is labeled at the upper left corner in each panel in units of km s$^{-1}$. Blue filled ellipses show the size of the synthesized beam.}\label{chant2}
\end{figure*}

Fig~\ref{chant2} presents the distributions of the test particles having different line-of-sight velocities projected on the plane of the sky in the kinematical model with $V_{\rm rel}$ of 0.6 km s$^{-1}$ starting with an axisymmetric initial particle distribution after evolution of 10$^4$ yr in comparison with the observed velocity channel maps. 
At $V_{\rm LSR}$ of 4.7--5.4 km s$^{-1}$, the particle distributions overlap with the blueshifted arc-like structure in the observations, suggesting that including the relative motion between \object{HL~Tau} and its protostellar envelope can explain the observed high-velocity in the blueshifted arc-like structure. 
Nevertheless, the particle distributions in the kinematical model do not show similar morphologies to the observed arc-like structures.
Those particles that do not overlap with the observed intensity distributions above the 3$\sigma$ level are removed from the initial particle distribution, resulting in the model envelopes with asymmetric initial density distributions shown in Fig.~\ref{initpar}.

\section{Synthetic images of different envelope models}\label{modimage}
\begin{figure*}
\centering
\includegraphics[width=17cm]{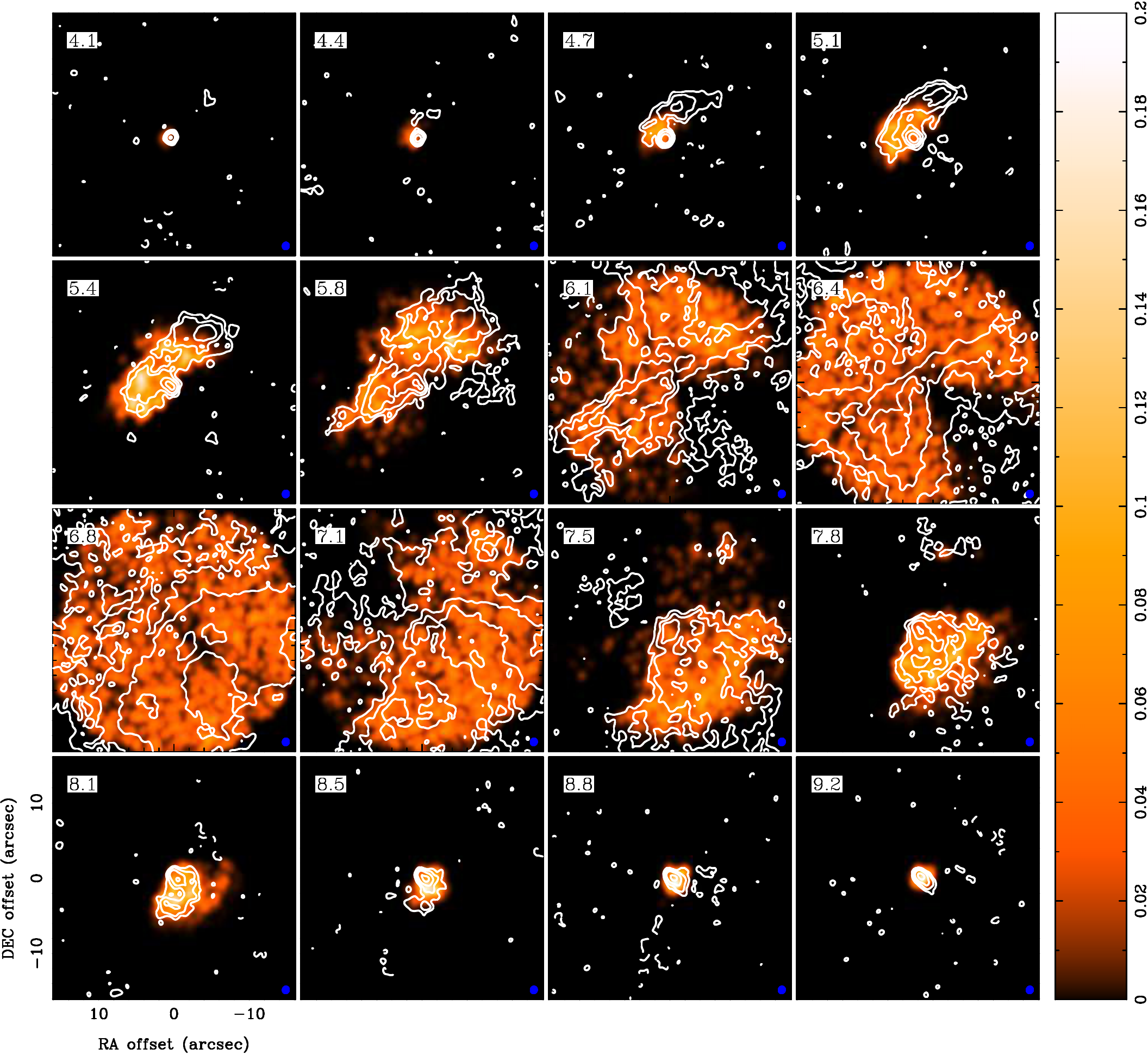}
\caption{Same as Fig.~\ref{mod0} but with $V_{\rm rel}$ of 0.3 km s$^{-1}$.}\label{mod03}
\end{figure*}

\begin{figure*}
\centering
\includegraphics[width=17cm]{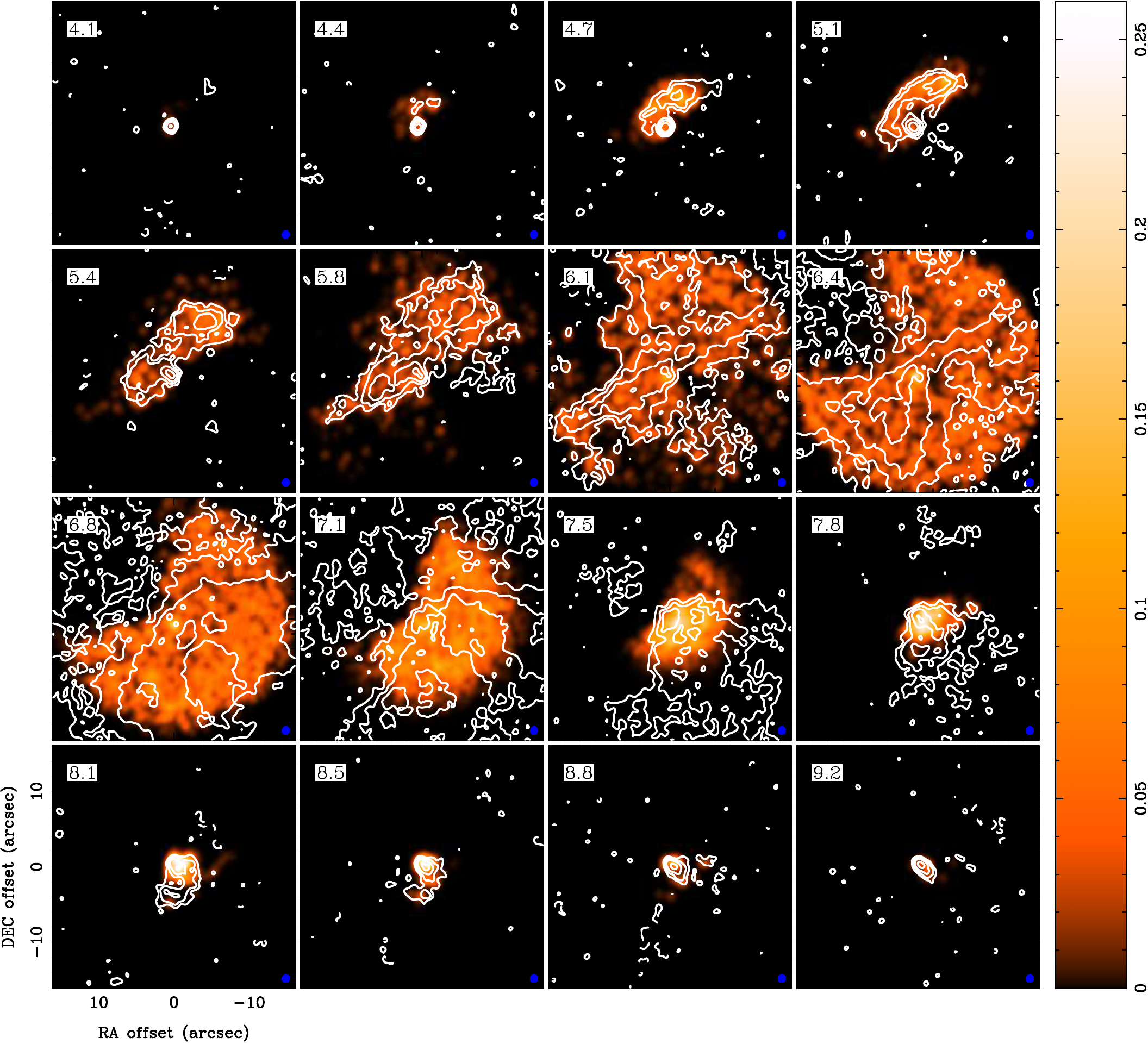}
\caption{Same as Fig.~\ref{mod0} but with $V_{\rm rel}$ of 0.9 km s$^{-1}$.}\label{mod09}
\end{figure*}

For comparison, Fig.~\ref{mod03} and \ref{mod09} present the synthetic velocity channel maps generated from the kinematical models with $V_{\rm vel}$ of 0.3 and 0.9 km s$^{-1}$.

\end{appendix}

\end{document}